\documentclass[journal]{IEEEtran}

\usepackage[utf8]{inputenc} 
\usepackage[T1]{fontenc}
\usepackage{url}
\usepackage{ifthen}
\usepackage{cite}
\usepackage[cmex10]{amsmath} 
\usepackage{amssymb}
\usepackage{xifthen,xparse}

\usepackage{hyperref}
\usepackage{float}
\usepackage{supertabular}

\usepackage{tikz,pgfplots,tkz-graph}
\pgfplotsset{compat=newest}
\pgfplotsset{plot coordinates/math parser=false}
\usetikzlibrary{decorations.markings,calc,positioning,matrix}

\allowdisplaybreaks

\newif\iffull

\newtheorem{theorem}{Theorem}
\newtheorem{corollary}[theorem]{Corollary}
\newtheorem{remark}[theorem]{Remark}
\newtheorem{lemma}[theorem]{Lemma}
\newtheorem{definition}[theorem]{Definition}

\newcommand{\ket}[1]{\left\lvert #1 \right\rangle}

\NewDocumentCommand\ketbra{+m+g}{%
  \IfNoValueTF{#2}
    {\left\lvert #1 \right\rangle \left\langle #1 \right\vert}
  {\left\lvert #1 \right\rangle \left\langle #2 \right\rvert}%
}
\NewDocumentCommand\braket{+m+g}{%
  \IfNoValueTF{#2}
    {\left\langle #1 \vert #1 \right\rangle}
  {\left\langle #1 \vert #2 \right\rangle}%
}

\newcommand{\vecnot}[1]{\underline{#1}}

\newcommand{\llbr}{[\![}
\newcommand{\rrbr}{]\!]}

\newcommand{\lX}{\bar{X}}
\newcommand{\lZ}{\bar{Z}}

\newcommand{\syminn}[2]{\langle #1, #2 \rangle_{\text{s}}}

\begin{document}
\title{Kerdock Codes Determine Unitary 2-Designs} 


\author{%
	\IEEEauthorblockN{Trung Can\IEEEauthorrefmark{1},
   					 Narayanan Rengaswamy\IEEEauthorrefmark{1},
                     Robert Calderbank,
                     and Henry D. Pfister}%
    \thanks{\IEEEauthorrefmark{1}These two authors contributed equally to this work. This work was presented in part at the 2019 IEEE International Symposium on Information Theory~\cite{Can-isit19}.}%
    \thanks{T. Can is currently with the
                     Department of Mathematics,
                     California Institute of Technology,
                     Pasadena, CA 91125, USA.
            Most of this work was conducted while he was with the
                     Department of Mathematics,
                     Duke University,
                     Durham, North Carolina 27708, USA.
                     Email: tcan@caltech.edu}%
    \thanks{N. Rengaswamy, R. Calderbank and H. D. Pfister are with the
                     Department of Electrical and Computer Engineering,
                     Duke University,
                     Durham, North Carolina 27708, USA.
                     Email: \{narayanan.rengaswamy, robert.calderbank, henry.pfister\}@duke.edu}
    \thanks{This work was supported in part by the National Science Foundation (NSF) under Grant Nos. 1718494 and 1908730. 
Any opinions, findings, conclusions, and recommendations expressed in this material are those of the authors and do not necessarily reflect the views of these sponsors.}
    \thanks{Copyright (c) 2017 IEEE. Personal use of this material is permitted.  However, permission to use this material for any other purposes must be obtained from the IEEE by sending a request to pubs-permissions@ieee.org.}                 
}

\maketitle

\begin{abstract}
The non-linear binary Kerdock codes are known to be Gray images of certain extended cyclic codes of length $N = 2^m$ over $\mathbb{Z}_4$.
We show that exponentiating these $\mathbb{Z}_4$-valued codewords by $\imath \triangleq \sqrt{-1}$ produces stabilizer states, that are quantum states obtained using only Clifford unitaries.
These states are also the common eigenvectors of commuting Hermitian matrices forming maximal commutative subgroups (MCS) of the Pauli group.
We use this \emph{quantum} description to simplify the derivation of the \emph{classical} weight distribution of Kerdock codes.
Next, we organize the stabilizer states to form $N+1$ mutually unbiased bases and prove that automorphisms of the Kerdock code permute their corresponding MCS, thereby forming a subgroup of the Clifford group.
When represented as symplectic matrices, this subgroup is isomorphic to the projective special linear group PSL($2,N$).
We show that this automorphism group acts transitively on the Pauli matrices, which implies that the ensemble is \emph{Pauli mixing} and hence forms a unitary $2$-design.
The \emph{Kerdock} design described here was originally discovered by Cleve et al. (2016), but the connection to classical codes is new which simplifies its description and translation to circuits significantly.
Sampling from the design is straightforward, the translation to circuits uses only Clifford gates, and the process does not require ancillary qubits. 
Finally, we also develop algorithms for optimizing the synthesis of unitary $2$-designs on encoded qubits, i.e., to construct \emph{logical} unitary $2$-designs.
Software implementations are available at \url{https://github.com/nrenga/symplectic-arxiv18a}, which we use to provide empirical gate complexities for up to $16$ qubits.
\end{abstract}


\section{Introduction}
\label{sec:intro}

\IEEEPARstart{Q}{uantum} computers promise enormous computational speedups over the best classical supercomputers in certain problems.
It has been established that even constant-depth quantum circuits provide an advantage over classical computation~\cite{Bravyi-science18}.
Today these devices are moving out of the lab and becoming generally programmable~\cite{Conover-sciencenews17}.
Since quantum computers are noisy, they will most likely employ quantum error-correcting codes (QECCs) to ensure reliability of computation.
\emph{Classical} error-correcting codes inspired the discovery of the first QECC by Shor~\cite{Shor-physreva95}, the development of CSS codes by Calderbank, Shor and Steane~\cite{Calderbank-physreva96,Steane-physreva96}, and the development of stabilizer codes~\cite{Calderbank-it98*2,Gottesman-phd97} by Calderbank, Rains, Shor and Sloane and by Gottesman.
A QECC protects $m-k$ logical qubits by embedding them in a system comprising $m$ physical qubits.
In fault-tolerant computation, any desired operation on the $m-k$ logical (protected) qubits must be implemented as a physical operation on the $m$ physical qubits that preserves the code space.
The most common QECCs are stabilizer QECCs, that are derived from commutative subgroups of the Heisenberg-Weyl group $HW_N$ (also known as the Pauli group), and the simplest quantum circuits are composed of unitary operators from the Clifford group $\text{Cliff}_N$ (see Section~\ref{sec:hw_clifford} for formal definitions).
Note that $N = 2^m$ throughout this paper.
The Clifford group normalizes $HW_N$, so each element of $\text{Cliff}_N$ induces an automorphism of $HW_N$ under conjugation.
Since $HW_N$ elements can be efficiently represented as binary length-$2m$ vectors, Clifford operators can be efficiently represented as $2m \times 2m$ binary \emph{symplectic} matrices that preserve the group structure of $HW_N$ under the binary mapping.
A central theme in this paper is that beyond inspiring the construction of QECCs, interactions between the classical and quantum worlds still prove mutually beneficial for other purposes, especially via the binary representations of $HW_N$ and $\text{Cliff}_N$.

Prior to running applications, a quantum computer needs to be assessed for its quality, in terms of the fidelity of the quantum operations executed on the system.
The \emph{randomized benchmarking} protocol, introduced by Emerson et al.~\cite{Emerson-joptb05,Magesan-physreva12}, is a well-established scheme that estimates the fidelity of the noise in the system by \emph{twirling} the underlying channel through a randomized sequence of gates and calculating the fidelity of the resulting depolarizing channel.
The depolarizing channel can be interpreted as the quantum analogue of the binary symmetric channel in classical communication theory.
Since the fidelity is shown to be an invariant of this twirling process, the protocol indeed estimates the fidelity of the actual noise (under some additional assumptions).
This protocol works on the physical operations in the computer, so the fidelity estimates then need to be translated into the quality of the \emph{logical} operations (on the protected qubits) for an error-corrected quantum computer. 
Since this translation might not be reliable, the procedure has recently been extended to the \emph{logical randomized benchmarking} protocol~\cite{Combes-arxiv17} that directly estimates the fidelity of the logical operations more reliably.

Randomized benchmarking requires that the sequence of gates be sampled from an ensemble of unitaries that form a \emph{unitary $2$-design}.
A unitary $t$-design is an ensemble of unitaries, bestowed with a probability distribution, that approximates the Haar distribution on the unitary group up to the $t$-th moment.
Section~\ref{sec:kerdock_twirl} makes this precise by discussing linear maps called \emph{twirls} over the full unitary ensemble and the finite ensemble.
When the two linear maps coincide the finite ensemble is a unitary $t$-design.
It is easy to analyze protocols that randomly sample unitary matrices with respect to the Haar measure, but such sampling is infeasible, hence the interest in finite ensembles of unitary matrices that approximate the Haar distribution.
$\text{Cliff}_N$ is known to be a unitary $3$-design~\cite{Webb-arxiv16}, and the proof by Webb involves the concepts of Pauli mixing and Pauli $2$-mixing, which we introduce in Section~\ref{sec:kerdock_twirl}.
However, $\text{Cliff}_N$ has size $2^{m^2 + 2m} \prod_{j=1}^{m} (4^j - 1)$ (up to scalars $e^{2\pi\imath \theta}, \theta \in \mathbb{R}$) which is much larger than the lower bound of $\approx 2^{4m}$ for any unitary $2$-design, established by Gross, Audenaert and Eisert~\cite{Gross-jmp07}, and by Roy and Scott~\cite{Roy-dcc09}.
They also discuss the existence of Clifford ensembles that saturate this bound.
Although random circuits are known to be exact or approximate unitary $2$-designs~\cite{Harrow-cmp09,Nakata-jmp17}, \emph{deterministic} constructions of such ensembles facilitate practical realizations.
Cleve et al.~\cite{Cleve-arxiv16} have recently found an explicit subgroup of $\text{Cliff}_N$ that is a unitary $2$-design.
The central contribution of this paper is to use the \emph{classical} Kerdock codes to simplify the construction of this \emph{quantum} unitary $2$-design and its translation to circuits (see Section~\ref{sec:kerdock_twirl}).

There are many other applications of unitary $2$-designs.
The first is quantum data hiding (the LOCC model described in~\cite{DiVincenzo-it02}), where the objective is to hide classical information from two parties who each share part of the data but are only allowed to perform local operations and classical communication.
The data hiding protocol can be implemented by sampling randomly from the full unitary group but it is sufficient to sample randomly from a unitary $2$-design.
Other applications of unitary $2$-designs are the fidelity estimation of quantum channels~\cite{Dankert-physreva09}, quantum state and process tomography~\cite{Scott-arxiv07}, and more recently minimax quantum state estimation~\cite{Quadeer-arxiv18}.
In quantum information theory, they have been used extensively in the analysis of decoupling~\cite{Hayden-arxiv07,Roy-dcc09,Szehr-njs13,Nakata-jmp17}.

The next section discusses the flow of ideas in the rest of the paper, and points to related work and applications.

\section{Main Ideas and Discussion}

The first result in the paper is to use the (quantum) commutative subgroups of $HW_N$ to simplify the derivation of weight distributions for the (classical) Kerdock codes. 

Recall that Hermitian matrices that commute can be simultaneously diagonalized.
The Kerdock and Delsarte-Goethals binary codes are unions of cosets of the first order Reed-Muller (RM) code $\text{RM}(1,m)$, the cosets are in one-to-one correspondence with symplectic forms, and the weight distribution of a coset is determined by the rank of the symplectic form (see~\cite[Chapter 15]{Macwilliams-1977} for more details).
Section~\ref{sec:dg_gf_construction} reviews the $\mathbb{Z}_4$ description of these codes given in~\cite{Hammons-it94}.
Section~\ref{sec:wt_dist} shows that when we exponentiate codewords in these cosets of $\text{RM}(1,m)$ we obtain a basis of common eigenvectors for a \emph{maximal} commutative subgroup of Hermitian Pauli matrices.
The operators that project onto individual lines in a given eigenbasis are invariants of the corresponding subgroup~\cite{Weyl-1997}, and the distance distribution of the code is determined by the inner product of pairs of such eigenvectors (see [Section~\ref{sec:ker_wt_dist}, Lemma~\ref{lem:herm_hamm}]).
When calculating weight distributions, this property makes it possible to avoid using Dickson's Theorem~\cite[Chapter 15]{Macwilliams-1977} to choose an appropriate representation of the symplectic form.
The correspondence between classical and quantum worlds simplifies the calculation (of the weight distribution) given in~\cite{Macwilliams-1977} significantly.
In order to demonstrate this simplicity, we provide the proof of the weight distribution here.
For most parts of the proof, we only need a brief description of the Kerdock set, Kerdock code, and the Gray map.
See Section~\ref{sec:ker_wt_dist} for formal definitions, constructions, and results used in the following proof.

In~\cite{Calderbank-arxiv10}, the \emph{Kerdock set} $P_{\text{K}}(m)$ is defined to be a collection of $N$ binary $m \times m$ symmetric matrices that is characterized by the following properties: it is closed under binary addition, and if $P,Q \in P_{\text{K}}(m)$ are distinct, then $P+Q$ is non-singular.
The codewords of the length-$N$ $\mathbb{Z}_4$-linear \emph{Kerdock code} $\text{K}(m)$ can be expressed as $[xPx^T + 2wx^T + \kappa]\ (\bmod\ 4)$, where $x \in \mathbb{F}_2^m$ chooses the symbol index in a codeword, and the codeword is determined by the choice of $w \in \mathbb{F}_2^m, P \in P_{\text{K}}(m)$ and $\kappa \in \mathbb{Z}_4 = \{0,1,2,3\}$.
One can obtain complex vectors of quaternary phases $\{1,\imath,-1,-\imath\}$ by raising $\imath = \sqrt{-1}$ to the integer powers in the ($\mathbb{Z}_4$-valued) codeword vector.
We will refer to this operation as exponentiating the codeword by $\imath$.
The \emph{Gray map} is an isometry from (length-$N$ vectors of quaternary phases, Euclidean metric) to ($\mathbb{Z}_2^{2N}$, Hamming metric) defined by $\mathfrak{g}(1) \triangleq 00, \mathfrak{g}(\imath) \triangleq 01, \mathfrak{g}(-1) \triangleq 11,$ and $\mathfrak{g}(-\imath) \triangleq 10$.
Note that the domain of the map can equivalently be taken as $\mathbb{Z}_4$ in which case the metric is the Lee metric (defined in Section~\ref{sec:wt_dist}).
The codewords of the \emph{binary non-linear} Kerdock code of length $2^{m+1}$ are obtained as Gray images of the $\mathbb{Z}_4$-valued codewords described above.

\begin{theorem} 
\label{thm:kerdock_wt_dist}
Let $m$ be odd. 
The weight distribution $A_i, i = 0,\ldots, 2^{m+1}$ of the classical binary Kerdock code of length $2^{m+1}$ is as follows.
\begin{table}[H]
\begin{center}
\begin{tabular}{c|c c c c c}
     
     $i$ & 0 & $2^m - 2^{(m-1)/2}$ & $2^m$ & $2^m + 2^{(m-1)/2}$ & $2^{m+1}$  \\ \hline \\ 
     $A_i$ & $1$ & $2^{2m + 1 } - 2^{m + 1}$ & $2^{m + 2} - 2$ & $2^{2m + 1} - 2^{m + 1}$ & $1$
     
\end{tabular}
\end{center}
\end{table}
\begin{IEEEproof}
We explicitly use Lemma~\ref{lem:herm_hamm} and Corollary~\ref{cor:ker_innerprods} to calculate the weight distribution.
Fix a vector $\mathsf{v}$ of quaternary phases obtained by exponentiating a codeword
\begin{align*}
c_{\mathsf{v}} = [{xP_2x^T + 2w_2x^T + \kappa_2}]_{x \in \mathbb{F}_2^m}
\end{align*}
in the $\mathbb{Z}_4$-linear Kerdock code.
Consider any vector $\mathsf{u}$ of quaternary phases obtained by exponentiating a second codeword
\begin{align*}
c_{\mathsf{u}} = [{xP_1x^T + 2w_1x^T + \kappa_1}]_{x \in \mathbb{F}_2^m}.
\end{align*}
(Lemmas~\ref{lem:eigenvecs} and~\ref{lem:eigenvec_innerprods} consider $\kappa_1 = \kappa_2 = 0$ so that the eigenvalues are $\pm 1$ (and not in $\{ \pm 1, \pm \imath \}$), but we account for these factors here.)

Let $n_j$ be the number of indices $x \in \mathbb{F}_2^m$ for which ${(c_{\mathsf{u}})_x - (c_{\mathsf{v}})_x = j}$, for $j \in \mathbb{Z}_4$.
Since the Gray map preserves the Lee metric, the Hamming distance between the Gray images of $c_{\mathsf{u}}$ and $c_{\mathsf{v}}$ is
$d_H(\mathfrak{g}(c_{\mathsf{u}}), \mathfrak{g}(c_{\mathsf{v}})) = n_1 + 2n_2 + n_3$.
Since $n_0 + n_1 + n_2 + n_3 = 2^m$ we simply have to relate $n_0$ and $n_2$ to obtain $d_H(\mathfrak{g}(c_{\mathsf{u}}), \mathfrak{g}(c_{\mathsf{v}}))$.
Observe that 
$\langle \mathsf{u}, \mathsf{v} \rangle = (n_0 - n_2) + (n_1 - n_3) \imath$.
Lemma~\ref{lem:herm_hamm} implies
\begin{align}
2^{m+1} - 2 \text{Re}[ \langle \mathsf{u}, \mathsf{v} \rangle ] 
  & = 2 d_H(\mathfrak{g}(c_{\mathsf{u}}), \mathfrak{g}(c_{\mathsf{v}})) \nonumber \\
\Rightarrow d_H(\mathfrak{g}(c_{\mathsf{u}}), \mathfrak{g}(c_{\mathsf{v}})) & = 2^m - (n_0 - n_2).
\end{align}
Now we observe three distinct cases for the codeword $c_{\mathsf{u}} - c_{\mathsf{v}}$.
Note that there are $2^{2m + 2}$ codewords in $\text{K}(m)$.
\begin{itemize}

\item[(i)] $P_1 = P_2, w_1 = w_2$: If $\kappa_1 - \kappa_2 = 0$ then we have the all-zeros codeword, and if $\kappa_1 - \kappa_2 = 2$ then we have the all-ones codeword.
However, if $\kappa_1 - \kappa_2 \in \{1,3\}$ then $n_0 - n_2 = 0$ and this determines two codewords of weight $2^m$ (more precisely, at distance $2^m$ from $c_{\mathsf{v}}$).

\item[(ii)] $P_1 = P_2, w_1 \neq w_2$: From Corollary~\ref{cor:ker_innerprods}, irrespective of $\kappa_1, \kappa_2$, we have $\langle \mathsf{u}, \mathsf{v} \rangle = 0$, which implies $n_0 - n_2 = 0$ and hence the distance is $2^m$.
This determines another $(2^m - 1) 2^2 = 2^{m+2} - 4$ codewords of weight $2^m$.

\item[(iii)] $P_1 \neq P_2$: From Corollary~\ref{cor:ker_innerprods} we have $|\langle \mathsf{u}, \mathsf{v} \rangle|^2 = 2^m$, which implies $(n_0 - n_2)^2 + (n_1 - n_3)^2 = 2^m$.
Since $m$ is odd, and $n_j$ are non-negative integers, direct calculation shows that this means $(n_0 - n_2)^2 = (n_1 - n_3)^2 = 2^{m-1}$ and therefore $n_0 - n_2 = \pm 2^{(m-1)/2}$.
More formally, since the Gaussian integers $\mathbb{Z}[\imath]$ are a unique factorization domain, we have $(n_0 - n_2) + (n_1 - n_3) \imath = {(\pm 1 \pm \imath) 2^{(m-1)/2}}$ and this gives weights $2^m \pm 2^{(m-1)/2}$.
Thus, we have $2^{2m+2} - 2^{m+2}$ codewords remaining and it is easy to see that the signs occur equally often. 
Hence there are $2^{2m+1} - 2^{m+1}$ codewords of each weight. \hfill \IEEEQEDhere

\end{itemize}
\end{IEEEproof}
\end{theorem}

Lemma~\ref{lem:eigenvecs} shows that the (normalized) length-$N$ vectors of quaternary phases obtained by exponentiating Kerdock, or more generally Delsarte-Goethals, codewords are common eigenvectors of maximal commutative subgroups of $HW_N$. 
These eigenvectors are called \emph{stabilizer states} in the quantum information literature~\cite{Dehaene-physreva03} (also see end of Section~\ref{sec:hw_clifford}).
They can equivalently be described as those quantum states obtainable by applying Clifford unitaries to the basis state $\ket{0}^{\otimes m} = e_0 \otimes \cdots \otimes e_0 = [1,0,0,\ldots,0]^T$~\cite{Aaronson-pra04}.
Clifford elements act by conjugation on $HW_N$, permuting the maximal commutative subgroups, and fixing the ensemble of stabilizer states (see Section~\ref{sec:mubs}).
The Clifford group is highly symmetric, and it approximates the full unitary group in a way that can be made precise by comparing irreducible representations~\cite{Koike-jalg87,Weyl-1997}.
Kueng and Gross~\cite{Kueng-arxiv15} have shown that the ensemble of stabilizer states is a complex projective $3$-design; given a polynomial of degree at most $3$, the integral over the $N$-sphere can be calculated by evaluating the polynomial at stabilizer states, and taking a finite sum.
Stabilizer states also find application as measurements in the important classical problem of phase retrieval, where an unknown vector is to be recovered from the magnitudes of the measurements (see Kueng, Zhu and Gross~\cite{Kueng-arxiv16}).
A third application is unsourced multiple access, where there is a large number of devices (messages) each of which transmits (is transmitted) infrequently.
This provides a model for machine-to-machine communication in the Internet-of-Things (IoT), including the special case of radio-frequency identification (RFID), as well as neighbor discovery in ad-hoc wireless networks.
Here, Thompson and Calderbank~\cite{Thompson-isit18} have shown that stabilizer states associated with Delsarte-Goethals codes support a fast algorithm for unsourced multiple access that scales to $2^{100}$ devices (arbitrary $100$-bit messages).

Section~\ref{sec:mubs} constructs the $N + 1$ eigenbases (of $N$ stabilizer states each) determined by the Kerdock code of length $N$ over $\mathbb{Z}_4$, and shows that the corresponding maximal commutative subgroups partition the non-identity Hermitian Pauli matrices.
The eigenbases are mutually unbiased, so that unit vectors $\mathsf{u}, \mathsf{v}$ in different eigenbases satisfy $|\langle \mathsf{u}, \mathsf{v} \rangle| = N^{-\frac{1}{2}}$, and hence each eigenbasis looks like noise to the other eigenbases.
The Kerdock ensemble of $N(N + 1)$ complex lines is extremal; Calderbank et al.~\cite{Calderbank-lms97} have shown that any collection of unit vectors for which pairwise inner products have absolute value $0$ or $N^{-\frac{1}{2}}$ has size at most $N^2 + N$, and that any extremal example must be a union of eigenbases.
The group of Clifford symmetries of this ensemble, represented as binary symplectic matrices, is shown to be isomorphic to the projective special linear group $\text{PSL}(2,N)$, and hence its size is $(N+1)N(N-1) = 2^{3m} - 2^m$.
We note that the Kerdock ensemble also appears in the work of Tirkkonen
et al.~\cite{Tirkkonen-isit17}.

Section~\ref{sec:kerdock_twirl} defines a graph $\mathbb{H}_N$ whose vertices are labeled by (scalar multiples of) all non-identity (Hermitian) Pauli matrices, where two matrices (vertices) are joined (by an edge) if and only if they commute.
This graph is shown to be \emph{strongly regular}; every vertex has the same degree, and the number of vertices joined to two given vertices depends only on whether the two vertices are joined or not joined.
The automorphism group of this graph is the group of binary symplectic matrices $\text{Sp}(2m,\mathbb{F}_2)$.
A subgroup of $\text{Cliff}_N$ containing $HW_N$ is proven to be Pauli mixing if it acts transitively on vertices, and Pauli $2$-mixing if it acts transitively on edges and on non-edges.
These properties imply that Pauli mixing ensembles are unitary $2$-designs and Pauli $2$-mixing ensembles are unitary $3$-designs~\cite{Webb-arxiv16}.
The Clifford symmetries of the Kerdock ensemble (of stabilizer states), again represented as symplectic matrices, are shown to be transitive on the vertices of $\mathbb{H}_N$ and hence a unitary $2$-design.
Since the Clifford symmetries include all Hermitian Paulis, in addition to PSL($2,N$), the size of the Kerdock unitary $2$-design is $N^5 - N^3 \approx 2^{5m}$, which almost saturates the bound by Gross et al.~\cite{Gross-jmp07} discussed above.
The next step is to translate these symmetry elements into circuits for, say, randomized benchmarking.

T. Can has developed an algorithm~\cite{Can-sen18} that factors a $2m \times 2m$ binary symplectic matrix into a product of at most $6$ elementary symplectic matrices of the type shown in Table~\ref{tab:std_symp}.
The target symplectic matrix maps the (Hadamard) dual basis $X_N = E([I_m \mid 0]), Z_N = E([0 \mid I_m])$ (see Section~\ref{sec:hw_clifford} for notation) to a dual basis $X_N', Z_N'$. 
Then, row and column operations by the elementary matrices return $X_N', Z_N'$ to the original pair $X_N, Z_N$ thereby producing a decomposition of the target symplectic matrix.

Section~\ref{sec:kerdock_twirl} uses this decomposition to simplify the translation of the Kerdock unitary $2$-design into circuits.
The elementary symplectic matrices appearing in the product can be related to the Bruhat decomposition of the symplectic group $\text{Sp}(2m,\mathbb{F}_2)$ (see~\cite{Maslov-it17}).
When the algorithm is run in reverse it produces a random Clifford matrix that can serve as an approximation to a random unitary matrix.
This is an instance of the subgroup algorithm~\cite{Diaconis-peis87} for generating uniform random variables.
The algorithm has complexity $O(m^3)$ and uses $O(m^2)$ random bits, which is order optimal given the order of the symplectic group $\text{Sp}(2m,\mathbb{F}_2)$ (cf.~\cite{Koenig-jmp14}).
We note that the problem of selecting a unitary matrix uniformly at random finds application in machine learning (see~\cite{Choromanski-arxiv16} and the references therein).
The algorithm developed by Can is similar to that developed by Jones, Osipov and Rokhlin~\cite{Jones-acha13} in that it alternates (partial) Hadamard matrices and diagonal matrices; the difference is that the unitary $3$-design property of the Clifford group~\cite{Webb-arxiv16} provides randomness guarantees.

Finally, Section~\ref{sec:logical_design} constructs \emph{logical} unitary $2$-designs that can be applied in the logical randomized benchmarking protocol of Combes et al.~\cite{Combes-arxiv17}.
In prior work~\cite{Rengaswamy-arxiv18}, we have developed a mathematical framework for synthesizing \emph{all} physical circuits that implement a logical Clifford operator (on the encoded qubits) for stabilizer codes (up to equivalence classes and ignoring stabilizer freedom).
Circuit synthesis is enabled by representing the desired physical Clifford operator as a $2m \times 2m$ binary symplectic matrix.
For an $\llbr m,m-k \rrbr$ stabilizer code, every logical Clifford operator is shown to have $2^{k(k+1)/2}$ symplectic solutions, and these are enumerated efficiently using symplectic transvections, thus enabling optimization with respect to a suitable metric.
See \url{https://github.com/nrenga/symplectic-arxiv18a} for implementations.

It is now well-known that different codes yield efficient (e.g., low-depth) implementations of different logical operators.
However, computing environments can change dynamically so that qubits or qubit links might have varying fidelity, and thus low-depth alone might not be desirable.
Under such circumstances it is necessary to leverage all degrees of freedom in implementing a logical operator, and a compiler might use the above framework for this purpose.
More generally, a compiler might usefully switch between several codes~\cite{Anderson-prl14} \emph{dynamically}, depending on the state of the system.
Then this algorithm enables the compiler to be able to determine logical operators for a code quickly depending on the user-input circuit (on the \emph{protected} qubits).

Section~\ref{sec:logical_design} provides a proof of concept implementation of the Kerdock unitary $2$-design on the protected (logical) qubits of the $\llbr 6,4,2 \rrbr$ CSS code using the above logical Clifford synthesis algorithm.
The logical randomized benchmarking protocol requires a unitary $2$-design on the logical qubits, and Combes et al. use the full Clifford group for this purpose, which is much larger than the Kerdock design as shown above.

In summary, the purpose of this paper is to emphasize that interactions between the classical and quantum domains still prove mutually beneficial, as much as they helped inspire the first QECC more than two decades back.
Specifically, we make four main \emph{theoretical} contributions:
\begin{enumerate}

\item Use of quantum concepts to simplify the calculation of classical weight distributions of several families of non-linear binary codes~\cite{Nordstrom-ic67,Kerdock-ic72,Preparata-ic68,Goethals-elet74,Goethals-ic76,Delsarte-jct75,Macwilliams-1977,Huffman-2003}.

\item Elementary description of symmetries of the Kerdock code, and the $N^2 + N$ stabilizer states determined by this code~\cite{Calderbank-arxiv10,Thompson-globalsip17,Tirkkonen-isit17,Thompson-isit18}.

\item Demonstration that the symmetry group of the Kerdock code is a unitary $2$-design and that sampling from it is straightforward. 
Introduction of elementary methods for translation to circuits without using ancillary qubits.

\item Provide a proof of concept construction for unitary $2$-designs on the logical qubits of a stabilizer code~\cite{Rengaswamy-arxiv18,Combes-arxiv17}.

\end{enumerate}

We also provide \emph{software} implementations of all algorithms, at \url{https://github.com/nrenga/symplectic-arxiv18a}.
Using this utility, we provide empirical estimates of the gate complexity for circuits obtained from the Kerdock design.
We believe this paves the way for employing this design in several applications, specifically in randomized benchmarking~\cite{Magesan-physreva12,Helsen-arxiv17}.

\section{The Heisenberg-Weyl and Clifford Groups}
\label{sec:hw_clifford}


Quantum error-correcting codes serve to protect qubits involved in quantum computation, and this section summarizes the mathematical framework introduced in~\cite{Calderbank-physreva96, Calderbank-it98*2, Gottesman-phd97, Gottesman-arxiv97}, and described more completely in~\cite{Gottesman-arxiv09} and~\cite{Rengaswamy-arxiv18}.
In this framework for fault-tolerant quantum computation, Clifford operators on the $N$-dimensional complex space afforded by $m$ qubits are represented as $2m \times 2m$ binary symplectic matrices.
This is an exponential reduction in size, and the symplectic matrices serve as a binary control plane for the quantum computer.

\begin{remark}
\label{rem:notation1}
Throughout the paper, we adopt the convention that all binary vectors are row vectors, and $\mathbb{Z}_4$-, real- or complex-valued vectors are column vectors, where $\mathbb{Z}_4$ is the ring of integers modulo $4$.
The values $\imath^{\kappa}$, where $\imath \triangleq \sqrt{-1}, \kappa \in \mathbb{Z}_4$, are called \emph{quaternary phases}.
\end{remark}

A single qubit is a $2$-dimensional Hilbert space, and a quantum state $\mathsf{v}$ is a superposition of the two states ${e_0 \triangleq [1,0]^T}, {e_1 \triangleq [0,1]^T}$ which form the \emph{computational basis}.
Thus $\mathsf{v} = \alpha e_0 + \beta e_1$, where $\alpha, \beta \in \mathbb{C}$ satisfy $|\alpha|^2 + |\beta^2| = 1$ as per the Born rule~\cite[Chapter 3]{Wilde-2013}.
The \emph{Pauli} matrices are
\begin{align}
I_2, \ 
X \triangleq 
\begin{bmatrix}
0 & 1 \\
1 & 0
\end{bmatrix} , \ 
Z \triangleq 
\begin{bmatrix}
1 & 0 \\
0 & -1
\end{bmatrix} , \  
Y \triangleq \imath XZ = 
\begin{bmatrix}
0 & -\imath \\
\imath & 0
\end{bmatrix},
\end{align} 
where $\imath \triangleq \sqrt{-1}$ and $I_2$ is the $2 \times 2$ identity matrix~\cite[Chapter 10]{Nielsen-2010}.
We may express an arbitrary \emph{pure} quantum state $\mathsf{v}$ as
\begin{align}
\label{eq:v_Paulis}
\mathsf{v} = ( \alpha_0 I_2 + \alpha_1 X + \imath \alpha_2 Z + \alpha_3 Y )\ e_0, \ \text{where}\ \alpha_i \in \mathbb{R}.
\end{align}
We describe $m$-qubit states by (linear combinations of) $m$-fold Kronecker products of computational basis states, or equivalently by $m$-fold Kronecker products of Pauli matrices.

Given row vectors $a,b \in \mathbb{F}_2^m$ define the $m$-fold Kronecker product 
\begin{align}
\label{eq:d_ab}
D(a,b) \triangleq X^{a_1} Z^{b_1} \otimes \cdots \otimes X^{a_m} Z^{b_m} \in \mathbb{U}_N,\ N \triangleq 2^m,
\end{align}
where $\mathbb{U}_N$ denotes the group of all $N \times N$ unitary operators.
The \emph{Heisenberg-Weyl group} $HW_N$ (also called the \emph{$m$-qubit Pauli group}) consists of all operators $\imath^{\kappa} D(a,b)$, where $\kappa \in \mathbb{Z}_4 \triangleq \{0,1,2,3\}$.
The order is $|HW_N| = 4N^2$ and the \emph{center} of this group is $\langle \imath I_N \rangle \triangleq \{ I_N, \imath I_N, -I_N, -\imath I_N \}$, where $I_N$ is the $N \times N$ identity matrix.
Multiplication in $HW_N$ satisfies the identity
\begin{align}
\label{eq:hw_commute}
D(a,b) D(a',b') & = (-1)^{a' b^T + b' a^T} D(a',b') D(a,b) .
\end{align}
The standard \emph{symplectic inner product} in $\mathbb{F}_2^{2m}$ is defined as 
\begin{align}
\syminn{[a,b]}{[a',b']} \triangleq a' b^T + b' a^T = [a,b]\ \Omega \ [a',b']^T ,
\end{align}
where the symplectic form $\Omega \triangleq 
\begin{bmatrix}
0 & I_m \\ 
I_m & 0
\end{bmatrix}$ (see~\cite{Calderbank-it98*2,Rengaswamy-arxiv18}).
Therefore, two operators $D(a,b)$ and $D(a',b')$ commute if and only if $\syminn{[a,b]}{[a',b']} = 0$.

The homomorphism $\gamma \colon HW_N \rightarrow \mathbb{F}_2^{2m}$ defined by
\begin{align}
\label{eq:gamma}
\gamma(\imath^{\kappa} D(a,b)) \triangleq [a,b] \ \forall \ \kappa \in \mathbb{Z}_4
\end{align}
has kernel $\langle \imath I_N \rangle$ and allows us to represent elements of $HW_N$ (up to multiplication by scalars) as binary vectors.


The \emph{Clifford group} $\text{Cliff}_N$ consists of all unitary matrices $g \in \mathbb{C}^{N \times N}$ for which $g D(a,b) g^{\dagger} \in HW_N$ for all $D(a,b) \in HW_N$, where $g^{\dagger}$ is the Hermitian transpose of $g$~\cite{Gottesman-arxiv09}. 
$\text{Cliff}_N$ is the \emph{normalizer} of $HW_N$ in the unitary group $\mathbb{U}_N$.
The Clifford group contains $HW_N$ and its size is $|\text{Cliff}_N| =  2^{m^2 + 2m} \prod_{j=1}^{m} (4^j - 1)$ (up to scalars $e^{2\pi\imath \theta}, \theta \in \mathbb{R}$)~\cite{Calderbank-it98*2}.
We regard operators in $\text{Cliff}_N$ as \emph{physical} operators acting on quantum states in $\mathbb{C}^N$, to be implemented by quantum circuits.
Every operator $g \in \text{Cliff}_N$ induces an automorphism of $HW_N$ by conjugation.
Note that the inner automorphisms induced by matrices in $HW_N$ preserve every conjugacy class $\{ \pm D(a,b) \}$ and $\{ \pm \imath D(a,b) \}$, because~\eqref{eq:hw_commute} implies that elements in $HW_N$ either commute or anti-commute.
Matrices $D(a,b)$ are symmetric or anti-symmetric according as $ab^T = 0$ or $1$, hence the matrix
\begin{align}
E(a,b) \triangleq \imath^{ab^T} D(a,b)
\end{align}
is Hermitian.
Note that $E(a,b)^2 = I_N$.
The automorphism induced by a Clifford element $g$ satisfies
\begin{align}
\label{eq:symp_action}
g E(a,b) g^{\dagger} = \pm E\left( [a,b] F_g \right), \ {\rm where} \ \ F_g = 
\begin{bmatrix}
A_g & B_g \\
C_g & D_g
\end{bmatrix} 
\end{align}
is a $2m \times 2m$ binary matrix that preserves symplectic inner products: $\syminn{[a,b] F_g}{[a',b'] F_g} = \syminn{[a,b]}{[a',b']}$.
Hence $F_g$ is called a \emph{binary symplectic matrix} and the symplectic property reduces to $F_g \Omega F_g^T = \Omega$, or equivalently 
\begin{align}
\label{eq:symp_conditions}
A_g B_g^T = B_g A_g^T, \ C_g D_g^T = D_g C_g^T, \ A_g D_g^T + B_g C_g^T = I_m .
\end{align}
(See~\cite{Gosson-2006} for an extensive discussion on general symplectic geometry and quantum mechanics.)
The symplectic property encodes the fact that the automorphism induced by $g$ must respect commutativity in $HW_N$.
Let $\text{Sp}(2m,\mathbb{F}_2)$ denote the group of symplectic $2m \times 2m$ matrices over $\mathbb{F}_2$.
The map $\phi \colon \text{Cliff}_N \rightarrow \text{Sp}(2m,\mathbb{F}_2)$ defined by
\begin{align}
\phi(g) \triangleq F_g 
\end{align}
is a homomorphism with kernel $HW_N$, and every Clifford operator projects onto a symplectic matrix $F_g$.
Thus, $HW_N$ is a normal subgroup of $\text{Cliff}_N$ and $\text{Cliff}_N/HW_N \cong \text{Sp}(2m,\mathbb{F}_2)$.
This implies that the size is $|\text{Sp}(2m,\mathbb{F}_2)| = 2^{m^2} \prod_{j=1}^{m} (4^j - 1)$ (also see~\cite{Calderbank-it98*2}).
Table~\ref{tab:std_symp} lists elementary symplectic transformations $F_g$, that generate the binary symplectic group $\text{Sp}(2m,\mathbb{F}_2)$, and the corresponding unitary automorphisms $g \in \text{Cliff}_N$, which together with $HW_N$ generate $\text{Cliff}_N$.
(See~\cite[Appendix I]{Rengaswamy-arxiv18} for a discussion on the Clifford gates and circuits corresponding to these transformations.)

\begin{table}
\caption[caption]{\label{tab:std_symp}A generating set of symplectic matrices and their corresponding unitary operators.\\ {\normalfont The number of $1$s in $Q$ and $P$ directly relates to number of gates involved in the circuit realizing the respective unitary operators (see~\cite[Appendix I]{Rengaswamy-arxiv18}). The $N$ coordinates are indexed by binary vectors $v \in \mathbb{F}_2^m$, and $e_v$ denotes the standard basis vector in $\mathbb{C}^N$ with an entry $1$ in position $v$ and all other entries $0$. Here $H_{2^t}$ denotes the Walsh-Hadamard matrix of size $2^t$, $U_t = {\rm diag}\left( I_t, 0_{m-t} \right)$ and $L_{m-t} = {\rm diag}\left( 0_t, I_{m-t} \right)$. }}
\centering
\vspace{-0.1cm}
\begin{tabular}{c|c}
Symplectic Matrix $F_g$ & Clifford Operator $g$\\
\hline
~ & ~ \\
$\Omega = \begin{bmatrix} 0 & I_m \\ I_m & 0 \end{bmatrix}$ & $H_N = H_2^{\otimes m}$ \\
~ & ~ \\
$L_Q = \begin{bmatrix} Q & 0 \\ 0 & Q^{-T} \end{bmatrix}$ & $\ell_Q: e_{v} \mapsto e_{v Q}$ \\
~ & ~ \\
$T_P = \begin{bmatrix} I_m & P \\ 0 & I_m \end{bmatrix}; P = P^T$ & $t_P\ =\ {\rm diag}\left( \imath^{v P v^T \bmod 4} \right)$ \\
~ & ~ \\
$G_t = \begin{bmatrix} L_{m-t} & U_t \\ U_t & L_{m-t} \end{bmatrix}$ & $g_t = H_{2^t} \otimes I_{2^{m-t}}$ \\
~ & ~ \\
\hline
\end{tabular}
\vspace{-0.4cm}
\end{table}

We use commutative subgroups of $HW_N$ to define resolutions of the identity.
A \emph{stabilizer} is a subgroup $S$ of $HW_N$ generated by commuting Hermitian matrices of the form $\pm E(a,b)$, with the additional property that if $E(a,b) \in S$ then $-E(a,b) \notin S$~\cite[Chapter 10]{Nielsen-2010}.
The operators $\frac{I_N \pm E(a,b)}{2}$ project onto the $\pm 1$ eigenspaces of $E(a,b)$, respectively.

\begin{remark}
Since all elements of $S$ are unitary, Hermitian and commute with each other, they can be diagonalized simultaneously with respect to a common orthonormal basis, and their eigenvalues are $\pm 1$ with algebraic multiplicity $N/2$.
We refer to such a basis as the \emph{common eigenbasis} or simply \emph{eigenspace} of the subgroup $S$, and to the subspace of eigenvectors with eigenvalue $+1$ as the $\emph{+1}$ \emph{eigenspace} of $S$.
\end{remark}

If the subgroup $S$ is generated by $E(a_i,b_i), i = 1,\ldots,k$, then the operator
\begin{align}
\frac{1}{2^k} \prod_{i=1}^{k} (I_N + E(a_i,b_i))
\end{align}
projects onto the $2^{m-k}$-dimensional subspace $V(S)$ fixed pointwise by $S$, i.e., the $+1$ eigenspace of $S$.
The subspace $V(S)$ is the \emph{stabilizer code} determined by $S$.
We use the notation $\llbr m,m-k \rrbr$ code to represent that $V(S)$ encodes $m-k$ \emph{logical} qubits into $m$ \emph{physical} qubits.

Let $\gamma(S)$ denote the subspace of $\mathbb{F}_2^{2m}$ formed by the binary representations of the elements of $S$ using the homomorphism $\gamma$ in~\eqref{eq:gamma}.
A generator matrix for $\gamma(S)$ is
\begin{align}
\label{eq:stabilizer_G}
G_S \triangleq [a_i, b_i]_{i = 1,\ldots,k} \in \mathbb{F}_2^{k \times 2m}\ \text{s.t.}\ G_S\ \Omega\ G_S^T = 0,
\end{align}
where $0$ is the $k \times k$ matrix with all entries zero.

Given a stabilizer $S$ with generators $E(a_i,b_i), i = 1,\ldots,k$, we can define $2^k$ subgroups $S_{\epsilon_1 \cdots \epsilon_k}$ where the index $(\epsilon_1 \cdots \epsilon_k)$ represents that $S_{\epsilon_1 \cdots \epsilon_k}$ is generated by $\epsilon_i E(a_i,b_i)$, for $\epsilon_i \in \{ \pm 1 \}$.
Note that the operator
\begin{align}
\label{eq:stab_projectors}
\Pi_{\epsilon_1 \cdots \epsilon_k} \triangleq \frac{1}{2^k} \prod_{i=1}^{k} (I_N + \epsilon_i E(a_i,b_i))
\end{align}
projects onto $V(S_{\epsilon_1 \cdots \epsilon_k})$, and that
\begin{align}
\sum_{(\epsilon_1, \ldots, \epsilon_k) \in \{ \pm 1 \}^k} \Pi_{\epsilon_1 \cdots \epsilon_k} = I_N.
\end{align}
Hence the subspaces $V(S_{\epsilon_1 \cdots \epsilon_k})$, or equivalently the subgroups $S_{\epsilon_1 \cdots \epsilon_k}$, provide a resolution of the identity, and elements (errors) in $HW_N$ simply permute these subspaces (under conjugation).

Given an $\llbr m,m-k \rrbr$ stabilizer code, it is possible to perform encoded quantum computation in any of the subspaces $V(S_{\epsilon_1 \cdots \epsilon_k})$ by synthesizing appropriate logical Clifford operators (see~\cite{Rengaswamy-arxiv18} for algorithms).
If we think of these subspaces as \emph{threads}, then a computation starts in one thread and jumps to another when an error (from $HW_N$) occurs.
Quantum error-correcting codes enable error control by identifying the jump that the computation has made.
Identification makes it possible to modify the computation in flight instead of returning to the initial subspace and restarting the computation.
The idea of tracing these threads is called as \emph{Pauli frame tracking} in the literature (see~\cite{Chamberland-quantum17} and references therein).

A stabilizer group $S$ defined by $k = m$ generators is called a \emph{maximal commutative subgroup} of $HW_N$ and $\gamma(S)$ is called a \emph{maximal isotropic subspace} of $\mathbb{F}_2^{2m}$.
The generator matrix $G_S$ has rank $m$ and can be row-reduced to $[0 \mid I_m]$ if $S = Z_N \triangleq \{ E(0,b) \colon b \in \mathbb{F}_2^m \}$, or to the form $[I_m \mid P]$ if $S$ is disjoint from $Z_N$.
We will denote these subgroups as $E([0 \mid I_m])$ and $E([I_m \mid P])$, respectively.
The condition $G_S \Omega G_S^T = 0$ implies $P = P^T$, and any element of $\gamma(S)$ can be expressed in the form $[a,aP]$ for some $a \in \mathbb{F}_2^m$.
Note that $E([I_m \mid 0]) = X_N \triangleq \{ E(a,0) \colon a \in \mathbb{F}_2^m \}$.
Since $\dim V(S) = 2^{m-m} = 1$, the subgroup $S$ fixes exactly one vector.
The $N$ eigenvectors in an orthonormal eigenbasis for $S$ are defined up to an overall phase and called \emph{stabilizer states}~\cite{Dehaene-physreva03,Aaronson-pra04}.
The number of non-zero entries in a stabilizer state is determined by the intersection of $S$ with $Z_N$~\cite{Calderbank-tr10}.

\section{Weight Distributions of Kerdock Codes}
\label{sec:ker_wt_dist}


Kerdock codes were first constructed as non-linear binary codes~\cite{Kerdock-ic72}, as was the Goethals code~\cite{Goethals-elet74} and the Delsarte-Goethals codes~\cite{Delsarte-jct75}.
In this section, we describe the Kerdock and Delsarte-Goethals codes as linear codes over $\mathbb{Z}_4$, the ring of integers modulo $4$.
These $\mathbb{Z}_4$-linear codes were constructed by Hammons et al.~\cite{Hammons-it94} as Hensel lifts of binary cyclic codes, and this description requires Galois rings.
The description given in Section~\ref{sec:dg_gf_construction} requires finite field arithmetic, but is entirely binary and follows~\cite{Calderbank-arxiv10}.
Our construction of unitary $2$-designs in Section~\ref{sec:kerdock_twirl} uses the matrices that are defined in Section~\ref{sec:dg_gf_construction}.
In Section~\ref{sec:wt_dist}, we make a connection between the Kerdock and Delsarte-Goethals codes and maximal commutative subgroups of $HW_N$ via stabilizer states, use this relation to compute inner products between stabilizer states, and hence calculate weight distributions of Kerdock codes.

\subsection{Kerdock and Delsarte-Goethals Sets~\cite{Calderbank-arxiv10}}
\label{sec:dg_gf_construction}

The finite field $\mathbb{F}_{2^m}$ is obtained from the binary field $\mathbb{F}_2$ by adjoining a root $\alpha$ of a primitive irreducible polynomial $p(x)$ of degree $m$~\cite{McEliece-1987}.
The elements of $\mathbb{F}_{2^m}$ are polynomials in $\alpha$ of degree at most $m-1$, with coefficients in $\mathbb{F}_2$, and we will identify the polynomial $z_0 + z_1 \alpha + \ldots + z_{m-1} \alpha^{m-1}$ with the binary (row) vector $[z_0,z_1,\ldots,z_{m-1}]$.


The \emph{Frobenius map} $f\colon \mathbb{F}_{2^m} \rightarrow \mathbb{F}_{2^m}$ is defined by $f(x) \triangleq x^2$, and the \emph{trace map} $\text{Tr}\colon \mathbb{F}_{2^m} \rightarrow \mathbb{F}_2$ is defined by
\begin{align}
\text{Tr}(x) \triangleq x + x^2 + \ldots + x^{2^{m-1}}.
\end{align}
Since $(x+y)^2 = x^2 + y^2$ for all $x,y \in \mathbb{F}_{2^m}$, the trace is linear over $\mathbb{F}_2$.
The trace inner product $\langle x,y \rangle_{\text{tr}} = \text{Tr}(xy)$ defines a symmetric bilinear form, so there exists a binary symmetric matrix $W$ for which $\text{Tr}(xy) = xWy^T$.
In fact
\begin{align}
W_{ij} = \text{Tr}\left( \alpha^i \alpha^j \right), \ i,j = 0,1,\ldots,m-1.
\end{align}
The matrix $W$ is non-singular since the trace inner product is non-degenerate (if $\text{Tr}(xz) = 0$ for all $z \in \mathbb{F}_{2^m}$ then $x = 0$).
Observe that $W$ is a Hankel matrix, since if $i+j = h+k$ then $\text{Tr}(\alpha^i \alpha^j) = \text{Tr}(\alpha^h \alpha^k)$.
The matrix $W$ can be interpreted as the primal-to-dual-basis conversion matrix for $\mathbb{F}_{2^m}$, with the primal basis being $\{ 1, \alpha, \alpha^2, \ldots, \alpha^{m-1} \}$ (see~\cite{Cleve-arxiv16}).

The Frobenius map $f(x) = x^2$ is linear over $\mathbb{F}_2$, so there exists a binary matrix $R$ for which $f(x) \equiv xR$.
Since
\begin{align}
f(x_0 + x_1 \alpha & + \ldots + x_{m-1} \alpha^{m-1}) \nonumber \\ 
                   & = x_0 + x_1 \alpha^2 + \ldots + x_{m-1} \alpha^{2(m-1)},
\end{align}
the rows of $R$ are the vectors representing the field elements $\alpha^{2i}, i = 0,\ldots,m-1$.
Note that square roots exist for all elements of $\mathbb{F}_{2^m}$ since $R$ is invertible.

We write multiplication by $z \in \mathbb{F}_{2^m}$ as a linear transformation $xz \equiv xA_z$.
For $z = 0, A_0 = 0$, and for $z = \alpha^i$ the matrix $A_z = A^i$ for $i = 0,1,\ldots,2^m - 2$, where $A$ is the matrix that represents multiplication by the primitive element $\alpha$.
The matrix $A$ is the \emph{companion matrix} of the primitive irreducible polynomial $p(x) = p_0 + p_1 x + \ldots + p_{m-1} x^{m-1} + x^m$ over the binary field.
Thus
\begin{align}
A \triangleq 
\begin{bmatrix}
0 & 1 & 0 & \cdots & 0 \\
0 & 0 & 1 & \cdots & 0 \\
  & \vdots  &   & \ddots & \vdots \\
0 & 0 & 0 & \cdots & 1 \\
p_0 & p_1 & p_2 & \cdots & p_{m-1} 
\end{bmatrix},
\end{align}
and we have chosen $A$ rather than $A^T$ as the companion matrix since we are representing field elements in $\mathbb{F}_{2^m}$ by row vectors (rather than column vectors).

\begin{lemma} \label{lem:f2mrelations}
The matrices $A_z, W$, and $R^i$, for $i \in [m]$, satisfy:
\begin{enumerate}

\item[(a)] $A_z A_x = A_x A_z = A_{xz}$;

\item[(b)] $A_x + A_z = A_{x+z}$;


\item[(c)] $A_z W = W A_z^T$;

\item[(d)] $R^i A_x^{2^i} = A_x R^i,\ R^i A_x^2 = A_x^{2^{1-i}} R^i$, and $R^{-i} A_x^{-2} = A_x^{-2^{1+i}} R^{-i}$;

\item[(e)] $R^i W = W (R^{-i})^T$ and $W^{-1} R^{-i} W = (R^i)^T$.

\end{enumerate}
\begin{IEEEproof}
%
%
%
%
%
%
%
%
Identities (a) through (d) follow directly from the arithmetic of $\mathbb{F}_{2^m}$.
Specifically, for (c), observe that 
\begin{align*}
(x A_z) W y^T = \text{Tr}((xz)y) = \text{Tr}(x(yz)) = x W (y A_z)^T,
\end{align*}
and (d) can be proven similarly.
To prove part (e) we observe $\text{Tr}(x) = \text{Tr}(x^2)$ and verify that for all $x,y \in \mathbb{F}_{2^m}$,
\begin{align*}
(x R^i) W y^T = \text{Tr}(x^{2^i} y) = \text{Tr}(x y^{2^{-i}}) = x W (R^{-i})^T y^T.  \tag*{\IEEEQEDhere}
\end{align*}
\end{IEEEproof}
\end{lemma}

\begin{definition}
For $0 \leq r \leq (m - 1)/2$ and for $\vecnot{z} = (z_0,z_1, \dots, z_r) \in \mathbb{F}_{2^m}^{r+1}$ define the bilinear form 
$\beta_{\vecnot{z},r}(x,y) \triangleq \text{Tr}[z_0 xy + z_1(x^2 y + xy^2) + \ldots + z_r(x^{2^r} y + x y^{2^r})]$.
Note that $\beta_{\vecnot{z},r}(x,y)$ is represented by the binary symmetric matrix
\begin{align}
\label{eq:bin_symm_mats}
P_{\vecnot{z},r} \triangleq A_{z_0} W + \sum_{i=1}^{r} [A_{z_i} W (R^i)^T + R^i W A_{z_i}^T].
\end{align}
The \emph{Delsarte-Goethals set} $P_{\text{DG}}(m,r)$ consists of all such matrices $P_{\vecnot{z},r}$.
The \emph{Kerdock set} $P_{\text{K}}(m) \triangleq P_{\text{DG}}(m,0)$ consists of all matrices $P_z \triangleq P_{\vecnot{z},0}$, where $\vecnot{z} = (z), z \in \mathbb{F}_{2^m}$.
\end{definition}

\begin{lemma}
The Delsarte-Goethals set $P_{\text{DG}}(m,r)$ is an $m(r+1)$-dimensional vector space of symmetric matrices.
If $z \neq 0$ then $\text{rank}(P_{\vecnot{z},r}) \geq m-2r$.
Matrices in the Kerdock set $P_{\text{K}}(m)$ are non-singular.
\begin{IEEEproof}
Closure under addition follows from part (b) of Lemma~\ref{lem:f2mrelations}.
Observe $\text{Tr}(x) = \text{Tr}(x^2) = \cdots = \text{Tr}(x^{1/2})$.
If $x$ is in the nullspace of $P_{\vecnot{z},r}$, i.e., using its vector representation $x P_{\vecnot{z},r} = 0$, then $\beta_{\vecnot{z},r}(x,y) = 0$ for all $y \in \mathbb{F}_{2^m}$ and we obtain 
\begin{align*}
0 & = \text{Tr}[ z_0 xy + z_1(x^2y+xy^2) + \ldots + z_r(x^{2^r}y+xy^{2^r})] \\
  & = \text{Tr}[ (z_0 x)^{2^r} y^{2^r} + (z_1^{2^r} x^{2^{r+1}} y^{2^r}+ (z_1 x)^{2^{r-1}} y^{2^r}) + \ldots \\
  & \hspace*{3.5cm} \ldots + (z_r^{2^r} x^{2^{2r}} y^{2^r} + (z_r x) y^{2^r})] \\
  & = \text{Tr}[ y^{2^r} ( (z_0 x)^{2^r} + (z_1^{2^r} x^{2^{r+1}} + (z_1 x)^{2^{r-1}}) + \ldots \\
  & \hspace*{3.5cm} \ldots + (z_r^{2^r} x^{2^{2r}} + z_r x) )].
\end{align*}
This holds for all $y$, so  $(z_0 x)^{2^r} + (z_1^{2^r} x^{2^{r+1}} + (z_1 x)^{2^{r-1}}) + \ldots + (z_r^{2^r} x^{2^{2r}} + z_r x)$ must be identically $0$, i.e., $x$ is a root of the polynomial. 
Since the polynomial has at most $2^{2r}$ roots, the nullspace of $P_{\vecnot{z},r}$ has dimension at most $2r$, which implies that $\text{rank}(P_{\vecnot{z},r}) \geq m-2r$.  
\end{IEEEproof}
\end{lemma}

\begin{remark}
\label{rem:symm_matrix}
Note that since the dimension of the vector space of all binary $m \times m$ symmetric matrices is $m(m+1)/2$, the set $P_{\text{DG}}(m,(m-1)/2)$ contains \emph{all} possible symmetric matrices.
For the remainder of this paper we represent a general symmetric matrix as simply $P$, thereby dropping the subscripts $\vecnot{z},r$ unless necessary.
We will continue to represent Kerdock matrices as $P_z$.
\end{remark}

%
%
%

\subsection{Delsarte-Goethals Codes and Weight Distributions} 
\label{sec:wt_dist}

Hammons et al.~\cite{Hammons-it94} showed that the classical nonlinear Kerdock and Delsarte-Goethals codes, defined by quadratic forms in \cite{Kerdock-ic72,Goethals-ic76}, are images of linear codes over $\mathbb{Z}_4$ under the Gray map. 
In this section, we begin by reviewing this construction using the Kerdock and Delsarte-Goethals sets of matrices, and demonstrate that exponentiating these $\mathbb{Z}_4$-valued codewords entry-wise by $\imath$ produces stabilizer states. 
For stabilizer states of $E([I_m \mid P_{\vecnot{z}_1,r}])$ and $E([I_m \mid P_{\vecnot{z}_2,r}])$,
we calculate their Hermitian inner products using the trace of certain projection operators, and show that the distribution of inner products depends only on $\text{rank}(P_{\vecnot{z}_1,r} + P_{\vecnot{z}_2,r})$.
Then, since $\text{rank}(P_z) \in \{ 0,m \}$ for all $P_z \in P_{\text{K}}(m)$, we compute the weight distribution of Kerdock codes
by relating these Hermitian inner products to the histogram of values in the difference between two $\mathbb{Z}_4$-valued codewords.
In order to calculate the weight distribution of Delsarte-Goethals codes, we would need to determine the distribution of ranks in the Delsarte-Geothals sets $P_{\text{DG}}(m,r)$.
While this question is straightforward for $P_{\text{K}}(m)$, it remains open for general $P_{\text{DG}}(m,r)$ and will be investigated in future work.

\begin{definition} 
The $\mathbb{Z}_4$-linear \emph{Delsarte-Goethals code} is given by
\begin{align}
\label{eq:dgmr_codes}
\text{DG}(m,r) & \triangleq \{ [{xPx^T + 2wx^T + \kappa}]_{x \in \mathbb{F}_2^m} \colon P \in P_{\text{DG}}(m,r), \nonumber \\
   & \hspace*{3.5cm} w \in \mathbb{F}_2^m, \kappa \in \mathbb{Z}_4 \}.
\end{align}
This code has size $2^{m(r+1) + m + 2}$ and is of \emph{type} $4^{m+1} 2^{mr}$~\cite[Section 12.1]{Huffman-2003}.
The \emph{Kerdock code} $\text{K}(m) \triangleq \text{DG}(m,0)$.
\end{definition}

Here the notation $[{xPx^T + 2wx^T + \kappa}]_{x \in \mathbb{F}_2^m}$ represents a $\mathbb{Z}_4$-valued column vector with each entry $xPx^T + 2wx^T + \kappa\ (\text{mod}\ 4)$ indexed by the vector $x \in \mathbb{F}_2^m$.

\begin{definition}
For $u, v \in \mathbb{Z}_4^N$ the \emph{Lee weight} of $u$ is defined as $w_L(u) \triangleq n_1(u) + 2 n_2(u) + n_3(u)$, where $n_{\kappa}(u)$ denotes the number of entries of $u$ with value $\kappa$, and the \emph{Lee distance} (between $u$ and $v$) is defined as $d_L(u,v) \triangleq w_L(u-v)$.
\end{definition}

\begin{figure}[h]
\begin{center}

\begin{tikzpicture}

\draw (0,0) circle (1.5);
\draw[fill=black] (0,1.5) circle (0.1);
\draw[fill=black] (0,-1.5) circle (0.1);
\draw[fill=black] (1.5,0) circle (0.1);
\draw[fill=black] (-1.5,0) circle (0.1);

\node at (0,1.8) {$\mathfrak{g}(1/\imath) = 01$};
\node at (0,-1.8) {$\mathfrak{g}(3/-\imath) = 10$};
\node at (2.6,0) {$\mathfrak{g}(0/1) = 00$};
\node[align=right] at (-2.85,0) {$\mathfrak{g}(2/-1) = 11$};

\end{tikzpicture}

\end{center}
\caption{\label{fig:gray}The Gray map assigning integers modulo $4$ or quaternary phases to binary pairs (that are length-$2$ row vectors).}
\end{figure}
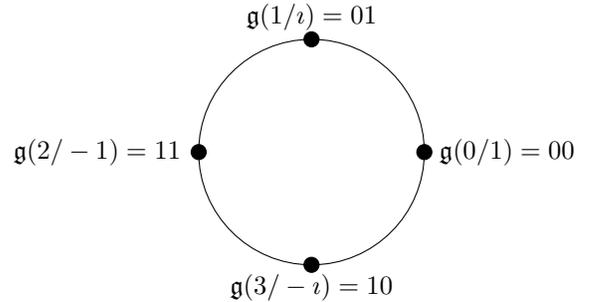

Figure~\ref{fig:gray} defines the \emph{Gray} map which assigns integers modulo $4$ (or quaternary phases) to binary pairs (see Remark~\ref{rem:notation1} for details).
For a vector, the map is applied to each entry and concatenated row-wise to return a row vector, thereby adhering to our convention for binary vectors (see Remark~\ref{rem:notation1}).
The shortest distance around the circle defines the Lee metric on $\mathbb{Z}_4^N$ and Gray encoding is an isometry from ($\mathbb{Z}_4^N$, Lee metric) to ($\mathbb{Z}_2^{2N}$, Hamming metric).
However, since $\mathfrak{g}(1 + 3) \neq \mathfrak{g}(1) + \mathfrak{g}(3)$, the Gray map is non-linear.
Hence the \emph{binary} Kerdock and Delsarte-Goethals codes obtained by Gray-mapping the codewords in $\text{K}(m)$ and $\text{DG}(m,r)$, respectively, are \emph{non-linear} (see~\cite[Chapter 12]{Huffman-2003}).

The Gray map is also a scaled isometry from (length-$N$ vectors of quaternary phases, squared Euclidean metric) to ($\mathbb{Z}_2^{2N}$, Hamming metric).
Note that for this set of quaternary phases, squared Euclidean distance is indeed a metric.
This is formalized in the following lemma.

\begin{lemma} 
\label{lem:herm_hamm}
Let $\mathsf{u},\mathsf{v} \in \mathbb{C}^{N}$ be two length-$N$ vectors of quaternary phases.
Then
\begin{align}
\langle \mathsf{u}-\mathsf{v}, \mathsf{u}-\mathsf{v} \rangle  = 2 d_H (\mathfrak{g}(\mathsf{u}),\mathfrak{g}(\mathsf{v})),
\end{align}
where $d_H$ denotes the Hamming distance.
\end{lemma}


Next, we prove a lemma establishing the relation between $\text{DG}(m,r)$ and the common eigenspace of $E([I_m \mid P])$ determined by a binary symmetric matrix $P$ (see Remark~\ref{rem:symm_matrix}).
Note that we denote the maximal commutative subgroup determined by the rows of $[I_m \mid P]$ as $E([I_m \mid P])$,
and that we do not normalize eigenvectors (stabilizer states) in this section, since the Gray map needs to be applied 
to quaternary phases.

\begin{lemma}
\label{lem:eigenvecs}
Given a binary symmetric matrix $P$, the (column) vectors $[\imath^{xPx^T + 2wx^T}]_{x \in \mathbb{F}_2^m}$ are common eigenvectors of the maximal commutative subgroup $E([I_m \mid P])$.
Each eigenvector has Euclidean length $\sqrt{N} = 2^{m/2}$.
\begin{IEEEproof}
%
It is possible to prove this result by direct calculation, i.e., by calculating $E(a,aP) \cdot [\imath^{xPx^T + 2wx^T}]_{x \in \mathbb{F}_2^m}$ for some $a \in \mathbb{F}_2^m$, but the following argument uses the mathematical framework described in Section~\ref{sec:hw_clifford}.
Note that 
\begin{align}
[\imath^{xPx^T + 2wx^T}]_{x \in \mathbb{F}_2^m} = \sum_{x \in \mathbb{F}_2^m} \imath^{xPx^T + 2wx^T} e_x,
\end{align}
where $e_x$ denotes the standard basis vector in $\mathbb{C}^N$ with a $1$ in the position $x$ and $0$ elsewhere.

The columns $[\imath^{2wx^T}]_{x \in \mathbb{F}_2^m} = [(-1)^{wx^T}]_{x \in \mathbb{F}_2^m}$ of the Walsh-Hadamard matrix $H_N$, where $w \in \mathbb{F}_2^m$ indexes the column, are common eigenvectors of the maximal commutative subgroup $X_N = E([I_m \mid 0])$.
The Clifford operator $t_P = \text{diag}\left( \imath^{xPx^T} \right)$ corresponds to the binary symplectic matrix $T_P = 
\begin{bmatrix}
I_m & P \\
0 & I_m
\end{bmatrix}$ (see Table~\ref{tab:std_symp}).
Hence conjugation by $t_P$ maps $E([I_m \mid 0])$ to $E([I_m \mid P])$, i.e., $t_P E(a,0) t_P^{\dagger} = E(a,aP)$, and so the common eigenvectors of $E([I_m \mid P])$ are 
\begin{align*}
t_P \cdot [\imath^{2wx^T}]_{x \in \mathbb{F}_2^m} = [\imath^{xPx^T + 2wx^T}]_{x \in \mathbb{F}_2^m}.   
\end{align*}
It is easily verified that for any $a \in \mathbb{F}_2^m$,
\begin{align*}
\left( t_P E(a,0) t_P^{\dagger} \right) \ t_P \cdot [\imath^{2wx^T}]_{x \in \mathbb{F}_2^m} = \pm [\imath^{xPx^T + 2wx^T}]_{x \in \mathbb{F}_2^m}.  \tag*{\IEEEQEDhere}
\end{align*}
\end{IEEEproof}
\end{lemma}

Now we have the following important observation.
%
%
%
%
%
%
%
Given $\mathsf{v} \in \mathbb{C}^N$, define
\begin{align}
S_{\mathsf{v}} \triangleq \{ g \in HW_N \colon g \mathsf{v} = \alpha \mathsf{v},\ \text{where}\ \alpha \in \mathbb{C}\ \text{and}\ |\alpha| = 1 \}. 
\end{align}

\begin{lemma}
\begin{enumerate}

\item[(a)] $S_{\mathsf{v}}$ is commutative.

\item[(b)] There is a $1$-$N$ correspondence between maximal commutative subgroups of $HW_N$ and stabilizer states.

\end{enumerate}
\begin{IEEEproof}
\begin{enumerate}

\item[(a)] If $E(a,b), E(a',b') \in S_{\mathsf{v}}$ then
\begin{align*}
E(a,b) E(a',b') \mathsf{v} = \alpha \alpha' \mathsf{v} = \alpha' \alpha \mathsf{v} = E(a',b') E(a,b) \mathsf{v}.
\end{align*}

\item[(b)] If $\mathsf{v}$ is a stabilizer state then $S_{\mathsf{v}}$ is a maximal commutative subgroup of $HW_N$. \hfill \IEEEQEDhere

\end{enumerate}
\end{IEEEproof}
\end{lemma}

\begin{remark}
\label{rem:dg_stab_states}
Any stabilizer state of a maximal commutative subgroup of $HW_N$ \emph{disjoint} from $Z_N$ can be obtained by exponentiating Delsarte-Goethals codewords, and multiplying the quaternary phase vector by $N^{-\frac{1}{2}}$.
The maximal commutative subgroups $E([I_m \mid P_z])$ determined by the Kerdock matrices $P_z$ intersect trivially.
Together with $Z_N = E([0 \mid I_m])$, they partition all $(N^2 - 1)$ non-identity Hermitian Pauli matrices.
Hence, given a non-identity Hermitian Pauli matrix $E(a,b)$, it follows that there is a sign $\epsilon \in \{ \pm 1 \}$ such that $\epsilon E(a,b)$ is in one of the $N+1$ subgroups determined by all $P_z \in P_{\text{K}}(m)$ and $Z_N$.
\end{remark}

Therefore, stabilizer states connect the \emph{classical} world of Kerdock and Delsarte-Goethals codes and the \emph{quantum} world of maximal commutative subgroups of $HW_N$.
This synergy has proven successful in several applications~\cite{Calderbank-arxiv10,Kueng-arxiv15,Kueng-arxiv16,Thompson-globalsip17,Tirkkonen-isit17,Thompson-isit18} and our construction of a unitary $2$-design here, from stabilizer states, is yet another instance.

\begin{remark}
Note that in Lemma~\ref{lem:eigenvecs} we only considered $\kappa = 0$ while exponentiating codewords $c_{\mathsf{u}} \in \text{DG}(m,r)$.
This is to ensure that the resulting eigenvector corresponds to a $\pm 1$ eigenvalue (and not a value in $\{ \pm 1, \pm \imath \}$).
However, Theorem~\ref{thm:kerdock_wt_dist} considers all $\kappa \in \mathbb{Z}_4$ while calculating the weight distribution.
\end{remark}


%
%
%
%
Given $P \in P_{\text{DG}}(m,r)$, we scale the common eigenvectors of $E([I_m \mid P])$ by $\sqrt{N}$ to obtain a set $V(P)$ of length-$N$ vectors of quaternary phases.
(Note the similarity to the notation $V(S)$ used in Section~\ref{sec:hw_clifford}, and observe that here we consider all eigenvectors, albeit unnormalized, and not just the $+1$ eigenspace.)
Therefore, if we can compute the Hermitian inner products between these unnormalized stabilizer states then we can use Lemma~\ref{lem:herm_hamm} to calculate the weight distribution of Kerdock and Delsarte-Goethals codes.
Note that despite being non-linear codes the weight and distance distributions of these codes coincide, as shown in~\cite[Chapter 15]{Macwilliams-1977} (which follows from $\mathbb{Z}_4$-linearity and Gray isometry).

\begin{lemma} 
\label{lem:eigenvec_innerprods}
Let $P_1, P_2 \in P_{\text{DG}}(m,r)$ be distinct. 
Fix $\mathsf{v} \in V(P_2)$ and let $\mathsf{u}$ run through $V(P_1)$. 
If $\text{rank}(P_1 + P_2) = k$ then 
\begin{align}
\lvert\langle \mathsf{u}, \mathsf{v} \rangle\rvert^2 = 
\begin{cases} 
2^{2m-k} & \text{for}\ 2^k\ \text{eigenvectors}\ \mathsf{u},\\ 
0  & \text{for}\ 2^m-2^k\ \text{eigenvectors}\ \mathsf{u} . 
\end{cases}
\end{align}
\begin{IEEEproof}
%
Let $Q = [I_m \mid P_1] \cap [I_m \mid P_2]$ represent a basis for the subspace formed by intersecting the spaces generated by $[I_m \mid P_1]$ and $[I_m \mid P_2]$.
Then $\dim(Q) = m-k$.
Let $[a_1,b_1], \ldots, [a_{m-k},b_{m-k}]$ be a basis for $Q$, and complete to bases for $P_1$ and $P_2$ by adding vectors $[c_j,d_j], j = m-k+1,\ldots,m$ and $[c_j',d_j'], j = m-k+1,\ldots,m$ respectively.
Since $\mathsf{v}$ is fixed, using~\eqref{eq:stab_projectors}, there are \emph{fixed} $f_i, t_j \in \{ \pm 1 \}$ such that
\begin{align*}
& \left( \frac{1}{\sqrt{N}} \mathsf{v} \right) \left( \frac{1}{\sqrt{N}} \mathsf{v}^{\dagger} \right) \\
& = \prod_{i=1}^{m-k} \frac{(I_N + f_i E(a_i,b_i))}{2} \prod_{j = m-k+1}^{m} \frac{(I_N + t_j E(c_j',d_j'))}{2},
\end{align*}
and since the only constraint for $\mathsf{u}$ is to be from $V(P_1)$, 
\begin{align*}
& \left( \frac{1}{\sqrt{N}} \mathsf{u} \right) \left( \frac{1}{\sqrt{N}} \mathsf{u}^{\dagger} \right) \\
& = \prod_{i=1}^{m-k} \frac{(I_N + e_i E(a_i,b_i))}{2} \prod_{j = m-k+1}^{m} \frac{(I_N + s_j E(c_j,d_j))}{2}, 
\end{align*}
where $e_i, s_i \in \{\pm 1\}$ are variable.
Since $\lvert \langle \mathsf{u}, \mathsf{v} \rangle \rvert^2 = \textrm{Tr}(\mathsf{u} \mathsf{u}^{\dagger} \mathsf{v} \mathsf{v}^{\dagger})$ 
it only remains to calculate $\text{Tr}(\mathsf{u} \mathsf{u}^{\dagger} \mathsf{v} \mathsf{v}^{\dagger})$.  
If $e_i = f_i \ \forall \ i$, then 
\begin{align*}
(I_N + e_i E(a_i,b_i)) (I_N + f_i E(a_i,b_i)) = 2(I_N + e_i E(a_i,b_i))
\end{align*}
so that
\begin{align*}
\text{Tr} (\mathsf{u} \mathsf{u}^{\dagger} \mathsf{v} \mathsf{v}^{\dagger}) & = 2^{m-k} \text{Tr} \biggr( \prod_{i=1}^{m-k}(I_N + e_i E(a_i,b_i)) \\
  & \hspace*{1.75cm} \times \prod_{j = m-k+1}^{m}(I_N + s_j E(c_j,d_j)) \\
  & \hspace*{2.5cm} \times \prod_{j = m-k+1}^{m}(I_N + t_j E(c_j',d_j')) \biggr).
\end{align*}
Expanding the right hand side, the only term with nonzero trace is the identity with trace $2^m$. 
Hence in this case $\text{Tr} (\mathsf{u} \mathsf{u}^{\dagger} \mathsf{v} \mathsf{v}^{\dagger}) = 2^{2m-k}$. 
The $k$ eigenvalues $s_j$ can be freely chosen, so there are $2^k$ eigenvectors in this case. 

If $e_i \neq f_i$ for some $i$, then 
\begin{align*}
(I_N + e_i E(a_i,b_i))(I_N + f_i E(a_i,b_i)) = 0
\end{align*}
and $\text{Tr}(\mathsf{u} \mathsf{u}^{\dagger} \mathsf{v} \mathsf{v}^{\dagger}) = 0$.
There are $2^m - 2^k$ such eigenvectors. 

Finally, if $k = m$ then $\text{Tr}(\mathsf{u} \mathsf{u}^{\dagger} \mathsf{v} \mathsf{v}^{\dagger}) = \text{Tr}(I_N) = 2^m \ \forall \ \mathsf{u}$.
\end{IEEEproof}
\end{lemma}

\begin{corollary}
\label{cor:ker_innerprods}
For $P_1, P_2 \in P_{\text{K}}(m)$, since $\text{rank}(P_1 + P_2) \in \{ 0,m \}$ the inner products are 
\begin{align}
\lvert\langle \mathsf{u}, \mathsf{v} \rangle\rvert^2 = 
\begin{cases} 
0  & \text{if}\ P_1 = P_2\ \text{and}\ \mathsf{u} \neq \mathsf{v} \\
2^m & \text{if}\ P_1 \neq P_2,\\
2^{2m} & \text{if}\ (P_1 = P_2\ \text{and})\ \mathsf{u} = \mathsf{v}. 
\end{cases}
\end{align}
\end{corollary}

This result is the primary tool that allowed us to simplify the derivation of the weight distribution of Kerdock codes in Theorem~\ref{thm:kerdock_wt_dist}.
Note that Theorem~\ref{thm:kerdock_wt_dist} is the only result in this section that is restricted to Kerdock codes but not general Delsarte-Goethals codes (and requires $m$ to be odd).

\section{Mutually Unbiased Bases from $P_{\text{K}}(m)$} 
\label{sec:mubs}


In this section, we will organize the columns of $I_N$ (i.e., the common eigenbasis of $Z_N$) and all stabilizer states determined by $P_z \in P_{\text{K}}(m)$ into a matrix to form mutually unbiased bases, and analyze its symmetries.
This symmetry group will eventually lead to the construction of the unitary $2$-design.
We first state a result that holds for stabilizer states determined by matrices from general Delsarte-Goethals sets. 

\begin{definition}
Given a collection $M$ of unit vectors in $\mathbb{C}^N$ (Grassmannian lines) the \emph{chordal distance} $\text{chor}(S)$ is given by
\begin{align}
\text{chor}(M) \triangleq \min_{(\mathsf{u}, \mathsf{v}) \in M} \sqrt{1 - |\langle \mathsf{u}, \mathsf{v} \rangle|^2}.
\end{align}
\end{definition}

It follows from Lemma~\ref{lem:eigenvec_innerprods} that the Delsarte-Goethals set $P_{\text{DG}}(m,r)$ determines $2^{m(r+2)}$ complex lines (stabilizer states) in $\mathbb{C}^N$ with chordal distance $\sqrt{1 - 2^{-(m-2r)}}$ (cf.~\cite{Calderbank-jac99}).

\begin{definition}
Two $N \times N$ unitary matrices $U$ and $V$ are said to be \emph{mutually unbiased} if $\lvert \langle \mathsf{u}, \mathsf{v} \rangle \rvert = N^{-\frac{1}{2}}$ for all columns $\mathsf{u}$ of $U$, and all columns $\mathsf{v}$ of $V$.
Each matrix is interpreted as an orthonormal basis and collections of such unitary matrices that are pairwise mutually unbiased are called \emph{mutually unbiased bases} (MUBs).
Vectors in each orthonormal basis look like noise to the other bases (due to the small inner product).
\end{definition}

Corollary~\ref{cor:ker_innerprods}, when applied to normalized eigenvectors, shows that the $N$ eigenbases determined by the Kerdock set $P_{\text{K}}(m)$ are mutually unbiased (also see Remark~\ref{rem:dg_stab_states}).
Let $\mathcal{B}_{\text{K}}(m)$ denote the collection of these $N$ eigenbases (of $E([I_m \mid P])$ for all $P \in P_{\text{K}}(m)$) along with the eigenbasis $I_N$ of $E([0 \mid I_m])$.
This is a set of $N+1$ mutually unbiased bases and they determine an ensemble of $N(N+1)$ complex lines (stabilizer states) that is extremal~\cite{Calderbank-lms97}. 
In this section, we provide an elementary description of their group of Clifford symmetries.

\subsection{The Kerdock MUBs}

Recollect from Section~\ref{sec:dg_gf_construction} that the Kerdock matrices are $P_z = A_z W$, where $W$ is a symmetric Hankel matrix with binary entries that satisfies $\text{Tr}(xy) = xWy^T$ and $A_z$ represents multiplication by $z$, both in $\mathbb{F}_{2^m}$.
Using the result of Lemma~\ref{lem:eigenvecs}, define $N$ mutually unbiased bases
\begin{align}
M_z \triangleq t_{P_z} H_N = \text{diag}\left( \imath^{xP_zx^T} \right) H_N,\ z \in \mathbb{F}_{2^m},
\end{align}
where $[H_N]_{x,y} \triangleq \frac{1}{\sqrt{N}} (-1)^{x y^T}$, for $x,y \in \mathbb{F}_2^m$, is the Walsh-Hadamard matrix of order $N$.
Note that $M_0 = H_N$ and that all columns of $M_z$ have Euclidean length $\frac{1}{\sqrt{N}}$.
Complete the MUBs by appending the matrix $M_{\infty} \triangleq I_N$.
The common eigenspaces of the maximal commutative subgroup ${E([I_m \mid P_z])}$ are the columns of $M_z$ and the common eigenspaces of $E([0 \mid I_m])$ are the standard unit coordinate vectors.
Hence the set of \emph{Kerdock MUBs}
\begin{align}
\label{eq:mub_set}
\mathcal{B}_{\text{K}}(m) \triangleq \{ I_N, M_z \colon z \in \mathbb{F}_{2^m} \}
\end{align} 
is a maximal collection of mutually unbiased bases~\cite{Calderbank-lms97}.


\subsection{Symmetries of Kerdock MUBs}

%
%

Let $M$ be the $N \times N(N+1)$ matrix given by
\begin{align}
M \triangleq \left[ \ M_{\infty} \ \mid \ M_0 \ \mid \ \cdots \ \mid \ M_z \ \mid  \ \cdots \ \right].
\end{align}
Note that $M_{\infty} = I_N$ and $M_0 = H_N$.

\begin{definition} 
A \emph{symmetry} of $M$ is a pair $(U,G)$ such that $UMG = M$, where $U$ is an $N \times N$ unitary matrix, and $G$ is a generalized permutation matrix, i.e., $G = \Pi D$ where $\Pi$ is a permutation matrix and $D$ is a diagonal matrix of quaternary phases.
\end{definition}

Observe that for any such symmetry, $G$ can undo the action of $U$ if and only if $U$ induces a (generalized) permutation on the columns of $M$.
Moreover, since $U$ is unitary it has to preserve inner products, so Corollary~\ref{cor:ker_innerprods} implies that $U$ can only permute the bases $M_z$ and permute columns within each basis, or equivalently permute the corresponding maximal commutative subgroups and permute elements within each subgroup, respectively, by conjugation.

\begin{lemma}
\label{lem:symm_clifford}
For any symmetry $(U,G)$ of $M$, the unitary matrix $U$ is an element of the Clifford group $\text{Cliff}_N$.
\begin{IEEEproof}
%
%
A Pauli matrix $E(a,b) \in E([I_m \mid P_z])$ that fixes $M_z$ can be written as $E(a,b) = \sum_{\mathsf{v} \in M_z} \epsilon_{\mathsf{v}} \mathsf{v} \mathsf{v}^{\dagger}$, where $\epsilon_{\mathsf{v}} = \pm 1$ for all $\mathsf{v}$.
Since $U$ permutes the eigenbases $M_z$, it follows that $U \mathsf{v} \in M_{z'}$, for some $z' \in \mathbb{F}_{2^m} \cup \{ \infty \}$, is fixed by $U E(a,b) U^{\dagger}$ which must again be a Pauli matrix.
Hence $U$ is a Clifford element.
\end{IEEEproof}
\end{lemma}

We first observe the symmetries induced by elements of $HW_N$.

\subsubsection{Pauli Matrices $E(a,0)$ and $E(0,b)$}
The group ${E([I_m \mid 0])}$ of Pauli matrices $E(a,0)$ fixes each column of $M_0 = H_N$ and acts transitively on the columns of each of the other $N$ blocks.
The group $E([0 \mid I_m])$ of Pauli matrices $E(0,b)$ fixes each column of $M_{\infty} = I_N$ and acts transitively on the columns of each of the remaining blocks.


\begin{definition} 
The \emph{projective special linear group} $\text{PSL}(2,2^m)$ is the group of all transformations 
\begin{align}
f(z) = \frac{az+b}{cz+d},\ \text{where}\ a,b,c,d \in \mathbb{F}_{2^m}, ad + bc = 1,
\end{align}
acting on the projective line $\mathbb{F}_{2^m} \cup \{ \infty \}$.
The transformation $f$ is associated with the action of the $2 \times 2$ matrix 
$\begin{bmatrix}
a & b \\
c & d
\end{bmatrix}$ on $1$-dimensional spaces, since
\begin{align*}
f \colon
\begin{bmatrix}
z \\ 1
\end{bmatrix} \mapsto 
\begin{bmatrix}
az + b \\ cz + d
\end{bmatrix} \equiv
 \begin{bmatrix}
1 \\ 0
\end{bmatrix} \ \text{or}\ 
\begin{bmatrix}
\frac{az+b}{cz+d} \\ 1
\end{bmatrix}.
\end{align*}
The \emph{projective linear group} $\text{P}\Gamma\text{L}(2,2^m)$ is the group of all transformations
\begin{align}
f(z) = \frac{az^{2^{-i}}+b}{cz^{2^{-i}}+d},\ \text{where}\ a,b,c,d \in \mathbb{F}_{2^m}, ad + bc = 1,
\end{align}
and $i \in \{ 0,1,\ldots,m-1 \}$.
The orders are 
\begin{align}
\lvert \text{PSL}(2,2^m) \rvert &= (N+1) N (N-1) = 2^{3m} - 2^m, \nonumber \\
\lvert \text{P}\Gamma\text{L}(2,2^m) \rvert &= (N+1) N (N-1) m.
\end{align}
\end{definition} 

Now we analyze the symmetries induced by elements of the binary symplectic group $\text{Sp}(2m,\mathbb{F}_2)$.

\subsubsection{Clifford Symmetries of $M$} \label{sec:cliff_symmetries}
The group $\text{PSL}(2,2^m)$ is generated by the transformations $z \mapsto z+x, z \mapsto zx$, and $z \mapsto 1/z$.
The group $\text{P}\Gamma\text{L}(2,2^m)$ is $\text{PSL}(2,2^m)$ enlarged by the Frobenius automorphisms $z \mapsto z^{2^{-i}} \equiv z R^{-i}$ discussed in Section~\ref{sec:dg_gf_construction}.
We realize each of these transformations as a symmetry of $M$.
We recall that $A_z W A_z^T = A_z^2 W$ from part (c) of Lemma~\ref{lem:f2mrelations}, and for convenience we work with maximal commutative subgroups $E([I_m \mid A_z^2 W])$, i.e., the Kerdock matrices are $P_z = A_z^2 W$.
Note that every field element $\beta \in \mathbb{F}_{2^m}$ is a square, so this is equivalent to $P_z = A_z W$.

\begin{enumerate}

\item[(i)] $z \mapsto z+x \ \text{becomes}\ [I_m \mid A_z^2 W] \mapsto [I_m \mid A_{x+z}^2 W] \colon$ 
\begin{align}
[I_m \mid A_z^2 W] 
\begin{bmatrix} 
I_m & A_x^2 W \\ 
0 & I_m 
\end{bmatrix} & = [I_m \mid (A_z^2 + A_x^2) W] \nonumber \\
\label{eq:Pk1}
              & \equiv [I_m \mid (A_{x+z}^2) W].
\end{align} 

\item[(ii)] $z \mapsto xz \ \text{becomes}\ [I_m \mid A_z^2 W] \mapsto [I_m \mid A_{xz}^2 W] \colon$
\begin{align}
[I_m \mid A_z^2 W] 
\begin{bmatrix} 
A_x^{-1} & 0 \\ 
0 & A_x^T 
\end{bmatrix} & = [A_x^{-1} \mid A_z^2 W A_x^T] \nonumber \\
              & = [A_x^{-1} \mid A_x A_z^2 W] \nonumber \\
\label{eq:Pk2}
              & \equiv [I_m \mid A_{xz}^2 W].
\end{align} 

\item[(iii)] $z \mapsto 1/z \ \text{becomes}\ [I_m \mid A_z^2 W] \mapsto [I_m \mid A_{z^{-1}}^2 W] \colon$
\begin{align}
[I_m \mid A_z^2 W] 
\begin{bmatrix} 
0 & I_m \\ 
I_m & 0 
\end{bmatrix} &
\begin{bmatrix} 
W^{-1} & 0 \\ 
0 & W^T
\end{bmatrix} \nonumber \\
  & = [A_z^2 W \mid I_m] 
\begin{bmatrix} 
W^{-1} & 0 \\ 
0 & W 
\end{bmatrix} \nonumber \\
  & = [A_z^2 \mid W] \nonumber \\
\label{eq:Pk3}
  & \equiv [I_m \mid A_{z^{-1}}^2 W].
\end{align} 
Note that if we start with $z = 0$, i.e., the subgroup $E([I_m \mid 0])$, then since $W$ is invertible the final subgroup is $E([0 \mid I_m])$, interpreted as $z = \infty$.
    
\item[(iv)] $z \mapsto z' \triangleq z^{2^{-i}} \ \text{becomes}\ [I_m \mid A_z^2 W] \mapsto [I_m \mid A_{z'}^2 W] \colon$
\begin{align}
[I_m \mid A_z^2 W] 
\begin{bmatrix} 
R^{-i} & 0 \\ 
0 & (R^i)^T 
\end{bmatrix} & = [R^{-i} \mid A_z^2 W (R^i)^T] \nonumber \\
              & = [R^{-i} \mid A_z^2 R^{-i} W] \nonumber \\
\label{eq:Pk4}
              & \equiv [I_m \mid A_{z'}^2 W].
\end{align} 

\end{enumerate}

%
Let $\mathfrak{P}_{\text{K},m}$ be the group of symplectic transformations generated by (i), (ii) and (iii) above, and let $\mathfrak{P}_{\text{K},m}^*$ be the group $\mathfrak{P}_{\text{K},m}$ enlarged by the generators (iv).
Thus, using notation in Table~\ref{tab:std_symp}, we have
\begin{align}
\label{eq:P_generators}
\mathfrak{P}_{\text{K},m} & \triangleq \langle T_{A_x^2 W},\ L_{A_x^{-1}},\ \Omega L_{W^{-1}} ; x \in \mathbb{F}_{2^m} \rangle \cong \text{PSL}(2,2^m), \\
\mathfrak{P}_{\text{K},m}^* & \triangleq \langle T_{A_x^2 W},\ L_{R^{-i} A_x^{-1}},\ \Omega L_{W^{-1}} ; x \in \mathbb{F}_{2^m} \rangle \cong \text{P}\Gamma\text{L}(2,2^m).
\end{align}
Although $T_W^2 = I_{2m}$, in the unitary group we have $t_W^2 = E(0,d_W)$. 
Therefore the corresponding Clifford elements will generate a group larger than $\text{PSL}(2,2^m)$ and $\text{P}\Gamma\text{L}(2,2^m)$.
Each symplectic matrix in the above groups can be transformed into a quantum circuit (or simply a unitary matrix) by expressing it as a product of standard symplectic matrices from Table~\ref{tab:std_symp} (see~\cite[Section II]{Rengaswamy-arxiv18}).

\begin{remark}
Note that $\Omega \notin \mathfrak{P}_{\text{K},m}$ but $\Omega L_{W^{-1}} \in \mathfrak{P}_{\text{K},m}$, which means $H_N$ does not permute columns of $M$ but $H_N \ell_{W^{-1}}$ does.
Hence, for example, to map $[0,a]$ to $[a,0]$ one sequence would be $(\Omega L_{W^{-1}}) \cdot L_{A_{b}^{-1}}$, where $b$ satisfies $a W^{-1} A_b^{-1} = a$.
This does not force $A_b^{-1} = W$.
\end{remark}

\begin{lemma}
\label{lem:P_structure}
Any element of $\mathfrak{P}_{\text{K},m}$ can be described as a product of at most $4$ basis symplectic matrices given in Section~\ref{sec:cliff_symmetries}.
\begin{IEEEproof}
Using the results in this Section, a general block permutation $[I_m \mid A_z^2 W] \mapsto [I_m \mid A_{\frac{az+b}{cz+d}}^2 W]$ is realized as
\begin{align*}
[I_m \mid A_z^2 W]
\begin{bmatrix}
A_d^2 & A_b^2 W \\
W^{-1} A_c^2 & (A_a^2)^T
\end{bmatrix}
& = [A_{cz+d}^2 \mid A_{az+b}^2 W] \\
& \equiv \left[ I_m \mid A_{\frac{az+b}{cz+d}}^2 W \right].
\end{align*}
It can be verified that the above is a valid symplectic matrix and satisfies all conditions in~\eqref{eq:symp_conditions}.
We now show that this matrix can be decomposed as a product of $4$ basis matrices.

We use the results in Lemma~\ref{lem:f2mrelations} and observe the following:
\begin{align*}
\begin{bmatrix} 
A_x^{-2} & 0 \\ 
0 & (A_x^2)^T
\end{bmatrix}
\begin{bmatrix} 
0 & W^T \\
W^{-1} & 0 
\end{bmatrix} = 
\begin{bmatrix} 
0 & A_x^{-2} W \\
W^{-1} A_x^2 & 0 
\end{bmatrix} ;
\end{align*}
\begin{align*}
\begin{bmatrix} 
I_m & A_y^2 W \\ 
0 & I_m 
\end{bmatrix}
\begin{bmatrix} 
0 & A_x^{-2} W \\
W^{-1} A_x^2 & 0 
\end{bmatrix} = 
\begin{bmatrix} 
A_{xy}^2 & A_x^{-2} W \\
W^{-1} A_x^2 & 0 
\end{bmatrix} ;
\end{align*}
\begin{align*}
\begin{bmatrix} 
A_{xy}^2 & A_x^{-2} W \\
W^{-1} A_x^2 & 0 
\end{bmatrix}
\begin{bmatrix} 
I_m & A_w^2 W \\ 
0 & I_m 
\end{bmatrix} = 
\begin{bmatrix} 
A_{xy}^2 & A_{wxy + x^{-1}}^{2} W \\
W^{-1} A_x^2 & (A_{wx}^2)^T
\end{bmatrix} .
\end{align*}
Hence we set $x \triangleq c, w \triangleq \frac{a}{c}, y \triangleq \frac{d}{c}$ so that $wxy + x^{-1} = \frac{ad+1}{c} = b$ and the resultant matrix matches the general symplectic matrix given above.
\end{IEEEproof}
\end{lemma}

\begin{corollary}
\label{cor:PSL_iso}
Let $a,b,c,d \in \mathbb{F}_{2^m}$ be such that $ad + bc = 1$. 
The isomorphism $\tau \colon \text{PSL}(2,2^m) \rightarrow \mathfrak{P}_{\text{K},m}$ can be defined as
\begin{align}
\tau \left(
\begin{bmatrix}
a & b \\
c & d
\end{bmatrix} \right) & \triangleq T_{A_{d/c}^2 W} \cdot L_{A_c^{-2}} \cdot \Omega L_{W^{-1}} \cdot T_{A_{a/c}^2 W}.
\end{align}
\end{corollary}

Observe that this provides a systematic procedure to sample from the group $\mathfrak{P}_{\text{K},m}$. 
By choosing $\alpha, \beta, \delta \in \mathbb{F}_{2^m}$ uniformly at random, a symmetry element can be constructed as
\begin{align}
\label{eq:sampleF}
F_{\alpha,\beta,\delta} \triangleq T_{A_{\alpha} W} \cdot L_{A_{\beta}} \cdot (\Omega L_{W^{-1}} \cdot T_{A_{\delta} W}).
\end{align}
The first two factors provide transitivity on the Hermitian matrices of all maximal commutative subgroups except $Z_N = E([0 \mid I_m])$, and the last factor enables exchanging any subgroup $E([I_m \mid P_z])$ with $E([0 \mid I_m])$ (see Lemma~\ref{lem:transitivity}).


We complete this section by observing that the symmetry group can be enlarged by including the Frobenius automorphisms $R$ from Section~\ref{sec:dg_gf_construction}.

\begin{lemma}
\label{lem:Pstar_structure}
An arbitrary element from $\mathfrak{P}_{\text{K},m}^*$ specified by $a,b,c,d \in \mathbb{F}_{2^m}$ and $i \in \{0,\ldots,m-1\}$ takes the form
\begin{align}
F = 
\begin{bmatrix}
R^{-i} A_d^2 & R^{-i} A_b^2 W \\
W^{-1} R^{-i} A_c^2 & (R^i)^T (A_a^2)^T
\end{bmatrix} ,
\end{align}
with $ad + bc = 1$, and realizes the block permutation
\begin{align}
[I_m \mid A_z^2 W] \longmapsto \left[ I_m \mid A_{\frac{az' + b}{cz' + d}}^2 W \right], \ z' \triangleq z^{2^{-i}}.
\end{align}
\begin{IEEEproof}
See Appendix~\ref{sec:enlarged_P}.
\end{IEEEproof}
\end{lemma}

\section{Unitary 2-Designs from the Kerdock MUBs}
\label{sec:kerdock_twirl}


In this section, we show that the unitary transformations determined by $\mathfrak{P}_{\text{K},m}$, along with Pauli matrices $D(a,b) \in HW_N$, form a unitary $2$-design.
We first define a graph on Pauli matrices, where Clifford elements act as graph automorphisms.
We then show that a group of automorphisms that acts transitively on vertices forms a unitary $2$-design.
Finally we show that a group of automorphisms that acts transitively on vertices, on edges, and on non-edges forms a unitary $3$-design.


\begin{definition}
The \emph{Heisenberg-Weyl graph} $\mathbb{H}_N$ has $N^2-1$ vertices, labeled by pairs $\pm E(a,b)$ with $[a,b] \neq [0,0]$ where vertices labeled $\pm E(a,b)$ and $\pm E(c,d)$ are joined if $E(a,b)$ commutes with $E(c,d)$.
We use $[a,b]$ to represent the vertex labeled $\pm E(a,b)$ and ${[a,b]\  \text{---}\ [c,d]}$ to represent an edge between two vertices.
\end{definition}

\begin{remark}
Elements of the Clifford group act by conjugation on $HW_N$, inducing automorphisms of the graph $\mathbb{H}_N$.
We shall distinguish two types of edges in $\mathbb{H}_N$.
\emph{Type $1$} edges connect vertices from the same maximal commutative subgroup $E([I_m \mid P_z]), z \in \mathbb{F}_{2^m}$, or from $E([0 \mid I_m])$.
\emph{Type $2$} edges connect vertices from different maximal commutative subgroups.
\end{remark}


To see that $\text{Aut}(\mathbb{H}_N) = \text{Sp}(2m,\mathbb{F}_2)$, determine a symplectic matrix to reduce an arbitrary graph automorphism to an automorphism $\pi$ that fixes $[e_i,0], [0,e_i], i = 1,\ldots,m$, then show that $\pi$ fixes every vertex.
This essentially amounts to solving for a symplectic matrix satisfying a linear system. 

\begin{definition}[{\hspace{1sp}\cite[Def. 2.4]{Cameron-1991}}]
A \emph{strongly regular graph} with parameters $(n,t,\lambda,\mu)$ is a graph with $n$ vertices, where each vertex has degree $t$, and where the number of vertices joined to a pair of distinct vertices $x, y$ is $\lambda$ or $\mu$ according as $x, y$ are joined or not joined respectively.
\end{definition}

\begin{lemma}
The Heisenberg-Weyl graph $\mathbb{H}_N$ is strongly regular with parameters
\begin{align}
n = N^2 - 1, \ t = \frac{N^2}{2} - 2, \ \lambda = \frac{N^2}{4} - 3, \ \mu = \frac{N^2}{4} - 1.
\end{align}
\begin{IEEEproof}
A vertex $[c,d]$ joined to a given vertex $[a,b]$ is a solution to $[a,b] \Omega [c,d]^T = 0$ (due to~\eqref{eq:hw_commute}).
This is a linear system with a single constraint, and after eliminating the solutions $[0,0]$ and $[a,b]$ we are left with $t = \frac{N^2}{2} - 2$ distinct vertices $[c,d]$ joined to $[a,b]$.

Given vertices $[a,b], [c,d]$ a vertex $[e,f]$ joined to both $[a,b]$ and $[c,d]$ is a solution to a linear system with two independent constraints.
When $[a,b]$ is not joined to $[c,d]$, we only need to eliminate the solution $[0,0]$.
When $[a,b]$ is joined to $[c,d]$ we need to eliminate $[0,0], [a,b]$ and $[c,d]$.
\end{IEEEproof}
\end{lemma}

\begin{remark}
The number of edges in $\mathbb{H}_N$ is $(N^2 - 1)(\frac{N^2}{2} - 2)$.
The number of type-$1$ edges is $(N+1)(N-1)(N-2)/2$ and the number of type-$2$ edges is $(N^2-1) (\frac{N^2}{2} - N)/2$.
\end{remark}

%

\begin{lemma}
\label{lem:transitivity}
\begin{enumerate}

\item[(a)] The symplectic group $\text{Sp}(2m,\mathbb{F}_2)$ acts transitively on vertices, on edges, and on non-edges of $\mathbb{H}_N$.

\item[(b)] The groups $\mathfrak{P}_{\text{K},m}$ and $\mathfrak{P}_{\text{K},m}^*$ act transitively on vertices of $\mathbb{H}_N$.

\end{enumerate}
\begin{IEEEproof}
Part (a) is well-known in symplectic geometry, and can also be proven by direct calculation using symplectic matrices.

(b) Since $\mathfrak{P}_{\text{K},m}$ acts transitively on maximal commutative subgroups $E([0 \mid I_m]), E([I_m \mid P_z]), z \in \mathbb{F}_{2^m}$ (see~\eqref{eq:Pk1} and~\eqref{eq:Pk3}), we need only show that $\mathfrak{P}_{\text{K},m}$ is transitive on a particular subgroup, say $E([I_m \mid 0])$.
If $a,b \in \mathbb{F}_{2^m}$ then there exists $c \in \mathbb{F}_{2^m}$ such that $b = ac$, and it follows from~\eqref{eq:Pk2} that the symplectic matrix 
$\begin{bmatrix}
A_c & 0 \\
0 & A_{c^{-1}}^T
\end{bmatrix}$ maps $[a,0]$ to $[b,0]$. \hfill \IEEEQEDhere
\end{IEEEproof}
\end{lemma}

\begin{remark}
The groups $\mathfrak{P}_{\text{K},m}$ and $\mathfrak{P}_{\text{K},m}^*$ are not transitive on edges of $\mathbb{H}_N$ because they cannot mix type-$1$ and type-$2$ edges.
\end{remark}


\begin{definition}[{\hspace{1sp}\cite{Webb-arxiv16}~\cite[Chap. 7]{Watrous-2018}}]
Let $k$ be a positive integer.
An ensemble $\mathcal{E} = \{ p_i, U_i \}_{i=1}^n$, where the unitary matrix $U_i$ is selected with probability $p_i$, is said to be a \emph{unitary $k$-design} if for all linear operators $X \in (\mathbb{C}^N)^{\otimes k}$
\begin{align}
\label{eq:channel_twirl}
\sum_{(p,U) \in \mathcal{E}} p\ U^{\otimes k} X (U^{\dagger})^{\otimes k} = \int_{\mathbb{U}_N} d\eta(U)\ U^{\otimes k} X (U^{\dagger})^{\otimes k},
\end{align}
where $\eta(\cdot)$ denotes the Haar measure on the unitary group $\mathbb{U}_N$.
The linear transformations determined by each side of~\eqref{eq:channel_twirl} are called \emph{$k$-fold twirls}.
A unitary $k$-design is defined by the property that the ensemble twirl coincides with the full unitary twirl.

We define the \emph{Kerdock twirl} to be the linear transformation of $(\mathbb{C}^N)^{\otimes 2}$ determined by the uniformly weighted ensemble consisting of $\phi^{-1}(\mathfrak{P}_{\text{K},m})$ along with Pauli matrices $D(a,b)$, where $\phi \colon \text{Cliff}_N/HW_N \rightarrow \text{Sp}(2m,\mathbb{F}_2)$ (from Section~\ref{sec:hw_clifford}).
Similarly, we define the $2$-fold action (in~\eqref{eq:channel_twirl}) of the ensemble determined by $\mathfrak{P}_{\text{K},m}^*$ as the \emph{enlarged Kerdock twirl}.
\end{definition}

\begin{definition}
An ensemble $\mathcal{E} = \{p_i, U_i\}_{i=1}^{n}$ of Clifford elements $U_i$ is \emph{Pauli mixing} if for every vertex $[a,b]$ the distribution $\{ p_i, U_i E(a,b) U_i^{\dagger} \}$ is uniform over vertices of $\mathbb{H}_N$.
The ensemble $\mathcal{E}$ is \emph{Pauli $2$-mixing} if it is Pauli mixing and if for every edge (resp. non-edge) $([a,b],[c,d])$ the distribution $\{ p_i, (U_i E(a,b) U_i^{\dagger}, U_i E(c,d) U_i^{\dagger}) \}$ is uniform over edges (resp. non-edges) of $\mathbb{H}_N$.
\end{definition}

\begin{theorem}
Let $G$ be a subgroup of the Clifford group containing all $D(a,b) \in HW_N$, and let $\mathcal{E} = \{ \frac{1}{|G|}, U \}_{U \in G}$ be the ensemble defined by the uniform distribution.
If $G$ acts transitively on vertices of $\mathbb{H}_N$ then $\mathcal{E}$ is a unitary $2$-design, and if $G$ acts transitively on vertices, edges and non-edges then $\mathcal{E}$ is a unitary $3$-design.
\begin{IEEEproof}
Transitivity means a single orbit so that random sampling from $G$ results in the uniform distribution on vertices, edges, or non-edges.
Hence transitivity on vertices implies $\mathcal{E}$ is Pauli mixing and transitivity on vertices, edges and non-edges implies $\mathcal{E}$ is Pauli $2$-mixing.
It now follows from~\cite{Webb-arxiv16} or~\cite{Cleve-arxiv16} that Pauli mixing (resp. Pauli $2$-mixing) implies $\mathcal{E}$ is a unitary $2$-design (resp. unitary $3$-design).
\end{IEEEproof}
\end{theorem}

%
%

\begin{corollary}
Random sampling from the Clifford group gives a unitary $3$-design and random sampling from the groups $\mathfrak{P}_{\text{K},m}, \mathfrak{P}_{\text{K},m}^*$ followed by a random Pauli matrix $D(a,b)$ gives unitary $2$-designs.
\end{corollary}

Pucha{\l}a and Miszczak have developed a useful Mathematica\textsuperscript{\textregistered} package \texttt{IntU}~\cite{Puchala-bpasts17} for symbolic integration with respect to the Haar measure on unitaries.
For small $m$, we used this utility to verify the equality in~\eqref{eq:channel_twirl} explicitly for the Kerdock $2$-design.

Sampling uniformly from the groups $\mathfrak{P}_{\text{K},m}$ can be achieved using the \emph{systematic} procedure shown in~\eqref{eq:sampleF}.
The resultant symplectic matrix can be transformed into a quantum circuit (or simply a unitary matrix) by expressing it as a product of standard symplectic matrices from Table~\ref{tab:std_symp} (see~\cite[Section II]{Rengaswamy-arxiv18}).
Although our unitary $2$-design is equivalent to that discovered by Cleve et al.~\cite{Cleve-arxiv16}, the methods we use to translate design elements to circuits are very different and much simpler.
While they use sophisticated methods from finite fields to propose a circuit translation that is \emph{tailored} for the design, our algorithm from~\cite{Can-sen18} (whose details are discussed in~\cite[Appendix II]{Rengaswamy-arxiv18}) is a \emph{general purpose} procedure that can be used to translate \emph{arbitrary} symplectic matrices to circuits.
They have been able to show Clifford-gate-complexity $O(m \log m \log \log m)$ assuming the extended Riemann hypothesis is true, or $O(m \log^2 m \log \log m)$ unconditionally, both of which are \emph{near-linear} when compared to the $O(m^2)$ gate-complexity for general Clifford elements.
Our open-source implementation\footnote{Implementations online: \url{https://github.com/nrenga/symplectic-arxiv18a}} allows one to construct the design for a specified number of qubits, and we use this utility to calculate \emph{worst-case} gate complexities on up to $16$ qubits.

\begin{figure}
\begin{center}

\scalebox{0.65}{%
%
%
\definecolor{mycolor1}{rgb}{0.00000,0.44700,0.74100}%
\definecolor{mycolor2}{rgb}{0.85000,0.32500,0.09800}%
\begin{tikzpicture}

\begin{axis}[%
width=4.521in,
height=3.566in,
at={(0.758in,0.481in)},
scale only axis,
xmin=0,
xmax=16,
ymin=0,
ymax=140,
axis background/.style={fill=white},
xmajorgrids,
ymajorgrids,
legend style={at={(0.269,0.82)}, anchor=south west, legend cell align=left, align=left, draw=white!15!black}
]
\addplot[ycomb, color=blue, thick, mark=o, mark options={solid, blue}] table[row sep=crcr] {%
1	1\\
2	2\\
3	4\\
4	6\\
5	9\\
6	14\\
7	18\\
8	25\\
9	29\\
10	37\\
11	46\\
12	57\\
13	67\\
14	78\\
15	90\\
16	103\\
};
\addplot[forget plot, color=white!15!black] table[row sep=crcr] {%
0	0\\
16	0\\
};
\addlegendentry{$T_{A_{\alpha} W}$ worst-case over all $\alpha$}

\addplot[very thick, color=red, dashed, mark=square, mark options={solid, thick, red}] table[row sep=crcr] {%
1	nan\\
2	0\\
3	3.15937885490117\\
4	8\\
5	14.1094665237117\\
6	21.2506151123568\\
7	29.265216037052\\
8	38.0391000173077\\
9	47.4855981423878\\
10	57.5364870807743\\
11	68.136568235267\\
12	79.2402055073958\\
13	90.8090026703979\\
14	102.810183529461\\
15	115.215425919366\\
16	128\\
};
\addplot[forget plot, color=white!15!black] table[row sep=crcr] {%
0	0\\
16	0\\
};
\addlegendentry{$m \log_2(m) \log_2(\log_2(m))$}

\end{axis}
\end{tikzpicture}%
}

\caption{\label{fig:T_Ac_P}The gate complexities for the element $T_{A_\alpha W}$ in~\eqref{eq:sampleF} for varying $m$.}
\end{center}
\end{figure}
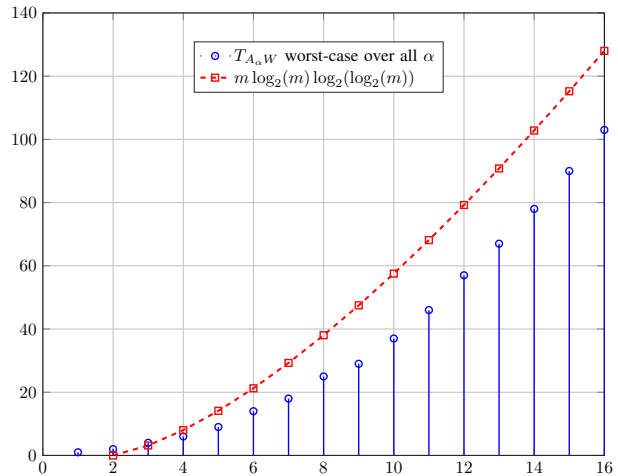

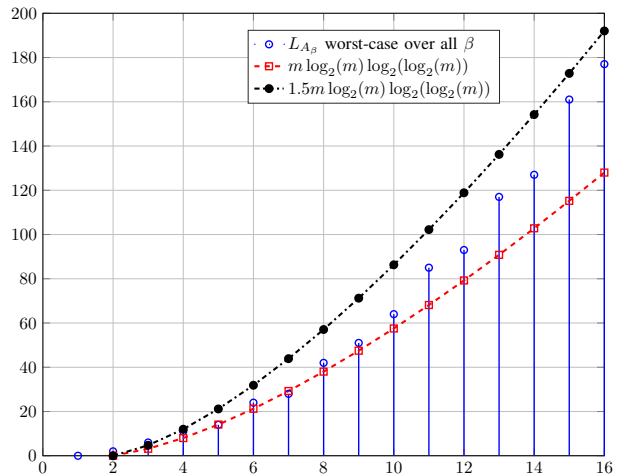
\begin{figure}
\begin{center}

\scalebox{0.65}{%
%
%
\definecolor{mycolor1}{rgb}{0.00000,0.44700,0.74100}%
\definecolor{mycolor2}{rgb}{0.85000,0.32500,0.09800}%
\definecolor{mycolor3}{rgb}{0.92900,0.69400,0.12500}%
\begin{tikzpicture}

\begin{axis}[%
width=4.521in,
height=3.566in,
at={(0.758in,0.481in)},
scale only axis,
xmin=0,
xmax=16,
ymin=0,
ymax=200,
axis background/.style={fill=white},
xmajorgrids,
ymajorgrids,
legend style={at={(0.365,0.796)}, anchor=south west, legend cell align=left, align=left, draw=white!15!black}
]
\addplot[ycomb, color=blue, thick, mark=o, mark options={solid, blue}] table[row sep=crcr] {%
1	0\\
2	2\\
3	6\\
4	11\\
5	14\\
6	24\\
7	28\\
8	42\\
9	51\\
10	64\\
11	85\\
12	93\\
13	117\\
14	127\\
15	161\\
16	177\\
};
\addplot[forget plot, color=white!15!black] table[row sep=crcr] {%
0	0\\
16	0\\
};
\addlegendentry{$L_{A_{\beta}}$ worst-case over all $\beta$}

\addplot[very thick, color=red, dashed, mark=square, mark options={solid, thick, red}] table[row sep=crcr] {%
1	nan\\
2	0\\
3	3.15937885490117\\
4	8\\
5	14.1094665237117\\
6	21.2506151123568\\
7	29.265216037052\\
8	38.0391000173077\\
9	47.4855981423878\\
10	57.5364870807743\\
11	68.136568235267\\
12	79.2402055073958\\
13	90.8090026703979\\
14	102.810183529461\\
15	115.215425919366\\
16	128\\
};
\addplot[forget plot, color=white!15!black] table[row sep=crcr] {%
0	0\\
16	0\\
};
\addlegendentry{$m \log_2(m) \log_2(\log_2(m))$}

\addplot[very thick, color=black, dashdotted, mark=*, mark options={solid, black}] table[row sep=crcr] {%
1	nan\\
2	0\\
3	4.73906828235175\\
4	12\\
5	21.1641997855676\\
6	31.8759226685352\\
7	43.897824055578\\
8	57.0586500259616\\
9	71.2283972135817\\
10	86.3047306211614\\
11	102.2048523529\\
12	118.860308261094\\
13	136.213504005597\\
14	154.215275294191\\
15	172.82313887905\\
16	192\\
};
\addplot[forget plot, color=white!15!black] table[row sep=crcr] {%
0	0\\
16	0\\
};
\addlegendentry{$1.5 m \log_2(m) \log_2(\log_2(m))$}

\end{axis}
\end{tikzpicture}%
}

\caption{\label{fig:L_Ac_inv}The gate complexities for the element $L_{A_\beta}$ in~\eqref{eq:sampleF} for varying $m$.}
\end{center}
\end{figure}

\begin{figure}
\begin{center}

\scalebox{0.65}{%
%
%
\definecolor{mycolor1}{rgb}{0.00000,0.44700,0.74100}%
\definecolor{mycolor2}{rgb}{0.85000,0.32500,0.09800}%
\begin{tikzpicture}

\begin{axis}[%
width=4.521in,
height=3.566in,
at={(0.758in,0.481in)},
scale only axis,
xmin=0,
xmax=16,
ymin=0,
ymax=140,
axis background/.style={fill=white},
xmajorgrids,
ymajorgrids,
legend style={at={(0.3,0.824)}, anchor=south west, legend cell align=left, align=left, draw=white!15!black}
]
\addplot[ycomb, color=blue, thick, mark=o, mark options={solid, blue}] table[row sep=crcr] {%
1	0\\
2	2\\
3	1\\
4	3\\
5	14\\
6	3\\
7	5\\
8	31\\
9	26\\
10	15\\
11	47\\
12	63\\
13	77\\
14	33\\
15	1\\
16	53\\
};
\addplot[forget plot, color=white!15!black] table[row sep=crcr] {%
0	0\\
16	0\\
};
\addlegendentry{$L_{W^{-1}}$}

\addplot[very thick, color=red, dashed, mark=square, mark options={solid, thick, red}] table[row sep=crcr] {%
1	nan\\
2	0\\
3	3.15937885490117\\
4	8\\
5	14.1094665237117\\
6	21.2506151123568\\
7	29.265216037052\\
8	38.0391000173077\\
9	47.4855981423878\\
10	57.5364870807743\\
11	68.136568235267\\
12	79.2402055073958\\
13	90.8090026703979\\
14	102.810183529461\\
15	115.215425919366\\
16	128\\
};
\addplot[forget plot, color=white!15!black] table[row sep=crcr] {%
0	0\\
16	0\\
};
\addlegendentry{$m \log_2(m) \log_2(\log_2(m))$}

\end{axis}
\end{tikzpicture}%
}

\caption{\label{fig:L_P_inv}The gate complexities for the element $L_{W^{-1}}$ in~\eqref{eq:sampleF} for varying $m$.}
\end{center}
\end{figure}
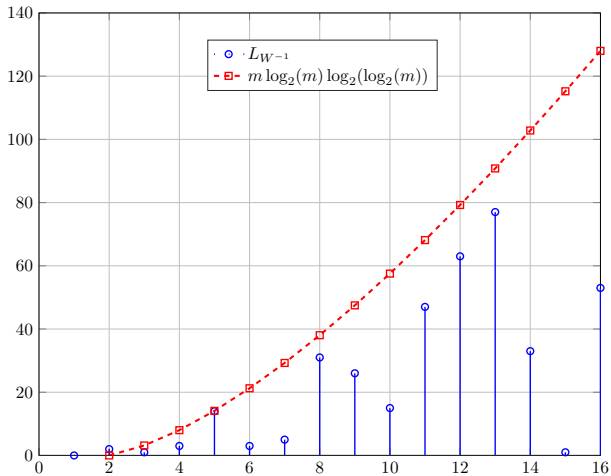

In our sampling procedure~\eqref{eq:sampleF} we have three elementary forms $T_{A_\alpha W}, L_{A_\beta}$, and $\Omega$, which translate to phase and controlled-$Z$ gates, permutations and controlled-NOT gates, and Hadamard gates on all qubits, respectively (see~\cite[Appendix I]{Rengaswamy-arxiv18}).
Note that $L_{W^{-1}}$ has the same elementary form as $L_{A_\beta}$, although $W$ is fixed for a given $m$.
The Hadamard gates add only $O(m)$ complexity.
Figures~\ref{fig:T_Ac_P},~\ref{fig:L_Ac_inv}, and~\ref{fig:L_P_inv} plot the \emph{worst-case} complexities of the gates $T_{A_\alpha W}, L_{A_\beta}$, and $L_{W^{-1}}$ obtained using our procedure.
The only form that seems to grow faster than $O(m \log m \log \log m)$ is $L_{A_{\beta}}$, and we are currently investigating methods to calculate this gate complexity via analytical arguments that leverage results in the classical computation literature.
A curious data point is $m = 15$ in Fig.~\ref{fig:L_P_inv}, where the matrix $W$ has zeros everywhere except the anti-diagonal, which translates to a single permutation of the qubits.
Since the decomposition in~\eqref{eq:sampleF} involves a \emph{constant} number of factors, the overall complexity is that of the factor with largest order term.
We will also investigate if our circuits can always be organized to give a depth of $O(\log m)$ just as Cleve et al.

Hence, we have provided an alternative perspective to the \emph{quantum} unitary $2$-design discovered by Cleve et al. by establishing a connection to \emph{classical} Kerdock codes, and simplified the description of the design as well as its translation to circuits.
Since we also appear to achieve competetive Clifford-gate-complexities, and provide implementations for our methods, we believe this paves the way for employing this $2$-design in several applications, specifically in randomized benchmarking~\cite{Magesan-physreva12,Helsen-arxiv17}.

\section{Logical Unitary $2$-Designs}
\label{sec:logical_design}


In this section, we apply our synthesis algorithm~\cite{Rengaswamy-arxiv18} to logical randomized benchmarking, which was recently proposed by Combes et al.~\cite{Combes-arxiv17} as a more precise protocol to reliably estimate performance metrics for an error correction implementation, as compared to the standard approach of physical randomized benchmarking.
Using this procedure, they are able to quantify the effects of imperfect logical gates, crosstalk, and correlated errors, which are typically ignored.
They use the full logical Clifford group to perform benchmarking as this group forms a (logical) unitary $2$-design (as well as $3$-design).
Our construction can be used to implement their protocol with a much smaller $2$-design ($\mathfrak{P}_{\text{K},m}$) and our synthesis algorithm can be used to realize the design at the logical level.

Here we show by example that we can efficiently translate \emph{any} unitary $2$-design of Clifford elements into a \emph{logical} unitary $2$-design, i.e., for a given $\llbr m, m-k \rrbr$ stabilizer code, produce $m$-qubit physical Clifford circuits that form a unitary $2$-design on the $m-k$ protected qubits.
As a proof of concept, we translate the Kerdock design $\mathfrak{P}_{\text{K},m-k}$ on the $m-k=4$ logical qubits of the $\llbr 6,4,2 \rrbr$ CSS code~\cite{Gottesman-arxiv97,Chao-arxiv17b} into their physical $6$-qubit implementations.
The stabilizers of this code are 
\begin{align}
S & \triangleq \langle X^{\otimes 6}, Z^{\otimes 6} \rangle \nonumber \\
  & = \langle E(111111,000000), E(000000,111111) \rangle.
\end{align}
The logical Pauli operators are 
\begin{align}
\begin{array}{c|c}
\lX_1 \triangleq E(110000,000000) & \lZ_1 \triangleq E(000000,010001) \\
\lX_2 \triangleq E(101000,000000) & \lZ_2 \triangleq E(000000,001001) \\
\lX_3 \triangleq E(100100,000000) & \lZ_3 \triangleq E(000000,000101) \\
\lX_4 \triangleq E(100010,000000) & \lZ_4 \triangleq E(000000,000011)
\end{array} .
\end{align}
Consider $\mathbb{F}_{16}$ constructed by adjoining to $\mathbb{F}_2$ a root $\alpha$ of the primitive polynomial $p(x) = x^4 + x + 1$.
Consider the element $F_{abcd}$ in $\mathfrak{P}_{\text{K},4}$ identified by the tuple $(a = \alpha^3, b = \alpha^8, c = \alpha^7, d = 0)$, or equivalently $(a,b,c,d,i=0)$ in $\mathfrak{P}_{\text{K},4}^*$.
In this case the matrices defined in Section~\ref{sec:dg_gf_construction} are 
\begin{align}
W = 
\begin{bmatrix}
0 & 0 & 0 & 1 \\
0 & 0 & 1 & 0 \\
0 & 1 & 0 & 0 \\
1 & 0 & 0 & 1
\end{bmatrix} & , 
W^{-1} = 
\begin{bmatrix}
1 & 0 & 0 & 1 \\
0 & 0 & 1 & 0 \\
0 & 1 & 0 & 0 \\
1 & 0 & 0 & 0
\end{bmatrix} , \nonumber \\
R = 
\begin{bmatrix}
1 & 0 & 0 & 0 \\
0 & 0 & 1 & 0 \\
1 & 1 & 0 & 0 \\
0 & 0 & 1 & 1
\end{bmatrix} & , \quad \ 
A = 
\begin{bmatrix}
0 & 1 & 0 & 0 \\
0 & 0 & 1 & 0 \\
0 & 0 & 0 & 1 \\
1 & 1 & 0 & 0
\end{bmatrix}.
\end{align}
Then, using the isomorphism in Corollary~\ref{cor:PSL_iso} and the direct form in Lemma~\ref{lem:P_structure}, we can express the element in $\mathfrak{P}_{\text{K},4}$ as
\begin{align}
F_{abcd} & = T_{A_{0}^2 W} \cdot L_{A_{\alpha^7}^{-2}} \cdot \Omega L_{W^{-1}} \cdot T_{A_{\alpha^{11}}^2 W} \nonumber \\
\label{eq:Fabcd1}
  & = I_{8} \cdot L_{(A^7)^{-2}} \cdot \Omega L_{W^{-1}} \cdot T_{(A^{11})^{2} W} \\
  & = 
\begin{bmatrix}
0 & A_{\alpha^8}^2 W \\
W^{-1} A_{\alpha^7}^2 & (A_{\alpha^3}^2)^T
\end{bmatrix} \nonumber \\
  & = 
\begin{bmatrix}
0 & A^{16} W \\
W^{-1} A^{14} & (A^{6})^T
\end{bmatrix} \\
\label{eq:Fabcd2}
  & = 
\left[
\begin{array}{cccc|cccc}
0 & 0 & 0 & 0 & 0 & 0 & 1 & 0 \\
0 & 0 & 0 & 0 & 0 & 1 & 0 & 0 \\
0 & 0 & 0 & 0 & 1 & 0 & 0 & 1 \\
0 & 0 & 0 & 0 & 0 & 0 & 1 & 1 \\
\hline
1 & 0 & 1 & 1 & 0 & 1 & 1 & 0 \\
0 & 1 & 0 & 0 & 0 & 1 & 0 & 1 \\
1 & 0 & 0 & 0 & 1 & 0 & 1 & 0 \\
1 & 0 & 0 & 1 & 1 & 1 & 0 & 1  
\end{array}
\right].
\end{align}
Using the explicit decomposition in~\eqref{eq:Fabcd1}, we map the elementary symplectic matrices to standard Clifford gates (as discussed in~\cite{Rengaswamy-arxiv18}) and get the following circuit (\texttt{CKT1}).\footnote{We list circuits instead of giving their circuit representation to align with the \texttt{MATLAB\textsuperscript{\textregistered}} cell array format we adopt in our implementations.}
\begin{center}
\begin{tabular}{ll}
`Permute' &	[4,1,2,3] \\ 
`CNOT'    & [1,2] \\
`Permute' & [4,3,2,1] \\
`CNOT'	  & [1,4] \\
`H'	      & [1,2,3,4] \\
`P'       &	[1,2,3,4] \\
`CZ'	  & [1,3] \\
`CZ'	  & [2,4] \\
`CZ'	  & [3,4]
\end{tabular}
\end{center}
Here the indices corresponding to `Permute' imply the cycle permutation $(1432)$, i.e., the first qubit has been replaced by the fourth, the fourth by the third, the third by the second, and the second by the first.
(Note that we do not simplify circuits to their optimal form here but simply report the results of our synthesis algorithm.)
An alternative procedure is to directly input the final symplectic matrix~\eqref{eq:Fabcd2} to the symplectic decomposition algorithm in~\cite{Can-sen18} (also see~\cite[Section II]{Rengaswamy-arxiv18}), yielding the following circuit (\texttt{CKT2}).
\begin{center}
\begin{tabular}{ll}
`Permute' & [3,2,1,4] \\
`CNOT'	  & [4,3] \\
`CNOT'	  & [1,4] \\
`H'	      & [1,2,3,4] \\
`P'	      & [1,2,3,4] \\
`CZ'	  & [1,3] \\
`CZ'	  & [2,4] \\
`CZ'	  & [3,4] 
\end{tabular}
\end{center}
The difference in depth of the two circuits is very small in this case, but we found that for about half of the elements in $\mathfrak{P}_{\text{K},4}$ the explicit form in Corollary~\ref{cor:PSL_iso} had smaller depth, while for those remaining, the direct decomposition was better.

Next we translate this logical circuit into its physical implementation for the $\llbr 6,4,2 \rrbr$ CSS code.
We apply our synthesis algorithm~\cite{Rengaswamy-arxiv18}, which can be summarized as follows.
\begin{enumerate}

\item Compute the action of either of the above circuits on the Pauli matrices $X_i, Z_i$, for $i = 1,2,3,4$, under conjugation, i.e., compute $g E(e_i,0) g^{\dagger}, g E(0,e_i) g^{\dagger}$ where $g$ represents the circuit.

\item Translate these into logical constraints on the desired physical circuit $\bar{g} \in \text{Cliff}_{2^6}$ by interpreting $X_i, Z_i$ above as their logical equivalents $\lX_i, \lZ_i \in HW_{2^6}$.

\item Rewrite the above conditions as linear constraints on the desired symplectic matrix $F_{\bar{g}}$ using~\eqref{eq:symp_action}.
Add constraints to normalize (or just centralize) the stabilizer $S$.

\item Solve for all symplectic solutions, compute their corresponding circuits and identify the best solution in terms of smallest depth, with respect to the decomposition in~\cite{Can-sen18} (also discussed in~\cite[Section II]{Rengaswamy-arxiv18}).

\item Verify the constraints imposed in step 2 and check for any sign violations (due to signs in~\eqref{eq:symp_action}). 
In case of violations, identify a Pauli matrix to fix the signs.

\end{enumerate}
Using this algorithm, we computed the circuit with smallest depth (relative to other solutions decomposed using the same algorithm) and obtained the following solution (\texttt{CKT3}).
\begin{center}
\begin{supertabular}{ll}
`Permute' & [1,6,3,5,4,2] \\
`CNOT'	  & [3,1] \\
`CNOT'	  & [4,1] \\
`CNOT'	  & [5,1] \\
`CNOT'	  & [6,1] \\
`CNOT'	  & [6,4] \\
`CNOT'	  & [6,5] \\
`CNOT'	  & [4,5] \\
`CNOT'	  & [3,6] \\
`CNOT'	  & [2,3] \\
`H'	      & [1,2,3,4,5,6] \\
`P'	      & [3,4,5] \\
`CZ'	  & [1,3] \\
`CZ'	  & [1,4] \\
`CZ'	  & [1,5] \\
`CZ'	  & [2,6] \\
`CZ'	  & [3,5] \\
`CZ'	  & [4,5] \\
`CZ'	  & [4,6] \\
`CZ'	  & [5,6] \\
`H'	      & [1,2] \\
`CNOT'	  & [2,3] \\
`CNOT'	  & [2,4] \\
`CNOT'	  & [2,5] \\
`CNOT'	  & [2,6] \\
`Z'	      & [2,6] \\
\end{supertabular}
\end{center}
We used the same procedure to translate all $4080$ elements of $\mathfrak{P}_{\text{K},4}$ into their (smallest depth) physical implementations for the $\llbr 6,4,2 \rrbr$ code, and the synthesis took about $25$ minutes on a laptop running the Windows 10 operating system (64-bit) with an Intel\textsuperscript{\textregistered} Core\textsuperscript{\texttrademark} i7-5500U @ 2.40GHz processor and 8GB RAM.
Note that for this case each element of $\mathfrak{P}_{\text{K},4}$ translates to $2^{k(k+1)/2} = 8$ symplectic solutions during the above procedure, and we need to compare depth after calculating circuits for all solutions.
In future work, we will try to optimize directly for depth without computing all solutions as this procedure is expensive for codes with large redundancy ($k$).
However, since this translation needs to be done only once for a given set $\mathfrak{P}_{\text{K},m-k}$ and an $\llbr m,m-k \rrbr$ stabilizer code, the circuits can be precomputed and stored in memory.

\section{Conclusion}
\label{sec:conclusion}

In this paper, we provided a simpler calculation of the Hamming weight distribution of the classical binary non-linear Kerdock codes by appealing to the connection with maximal commutative subgroups of the Heisenberg-Weyl group from the quantum domain.
Using this connection, we also described the group of Clifford symmetries of the mutually unbiased bases determined by the Kerdock sets of symmetric matrices.
We then used this result in the classical domain to construct small unitary $2$-designs.
Finally, we demonstrated an efficient procedure for translating any unitary $2$-design consisting of Clifford elements into a logical unitary $2$-design for a given stabilizer code.
This work demonstrates yet again that interactions between the classical and quantum domains are still mutually beneficial, both theoretically and practically.





\appendices

\section{Proof of Lemma~\ref{lem:Pstar_structure}}
\label{sec:enlarged_P}

We proceed as in the proof of Lemma~\ref{lem:P_structure} to derive the general form of an element in $\mathfrak{P}_{\text{K},m}^*$.
Introducing the new generators $L_{R^{-i}}$, and using identities from Lemma~\ref{lem:f2mrelations}, we calculate
\begin{align*}
& \begin{bmatrix} 
R^{-i} A_x^{-2} & 0 \\ 
0 & (R^i)^T (A_x^2)^T
\end{bmatrix} 
\begin{bmatrix} 
0 & W^T \\
W^{-1} & 0 
\end{bmatrix} \\
& = \begin{bmatrix} 
0 & R^{-i} A_x^{-2} W \\
(R^i)^T W^{-1} A_x^2 & 0 
\end{bmatrix} ; \\
& \begin{bmatrix} 
I_m & A_y^2 W \\ 
0 & I_m 
\end{bmatrix}
\begin{bmatrix} 
0 & R^{-i} A_x^{-2} W \\
(R^i)^T W^{-1} A_x^2 & 0 
\end{bmatrix} \\
& = \begin{bmatrix} 
A_{y}^2 W (R^i)^T W^{-1} A_x^2 & R^{-i} A_x^{-2} W \\
(R^i)^T W^{-1} A_x^2 & 0 
\end{bmatrix} ; \\
& \begin{bmatrix} 
A_{y}^2 W (R^i)^T W^{-1} A_x^2 & R^{-i} A_x^{-2} W \\
(R^i)^T W^{-1} A_x^2 & 0 
\end{bmatrix} 
\begin{bmatrix} 
I_m & A_w^2 W \\ 
0 & I_m 
\end{bmatrix} \\
& = \begin{bmatrix} 
A_{y}^2 (W (R^i)^T W^{-1}) A_x^2 & A_{y}^2 (W (R^i)^T W^{-1}) A_x^2 A_w^2 W + R^{-i} A_x^{-2} W \\
(R^i)^T W^{-1} A_x^2 & (R^i)^T W^{-1} A_x^2 A_w^2 W
\end{bmatrix} \\
& = \begin{bmatrix} 
A_{y}^2 R^{-i} A_x^2 & A_{y}^2 R^{-i} A_{wx}^2 W + R^{-i} A_x^{-2} W \\
W^{-1} R^{-i} A_x^2 & (R^i)^T (A_{wx}^2)^T
\end{bmatrix} \\
& \triangleq F.
\end{align*}
In this case the relations between $a,b,c,d$ and $x,y,z,w$ that will yield the desired map are unclear.
Hence, we first determine the transformation on ${[I_m \mid A_z^2 W]}$ in terms of $x,y,z$ and $w$.
Again, we repeatedly invoke identities from Lemma~\ref{lem:f2mrelations}.
\begin{align*}
& [I_m \mid A_z^2 W] F \\
& = [ A_{y}^2 R^{-i} A_x^2 + A_z^2 R^{-i} A_x^2 \mid (A_{y}^2 R^{-i} A_{wx}^2 + R^{-i} A_x^{-2}) W \\
& \hspace*{5.5cm} + A_z^2 W (R^i)^T (A_{wx}^2)^T ] \\
& = [ A_{y+z}^2 R^{-i} A_x^2 \mid (A_{y+z}^2 R^{-i} A_{wx}^2 + R^{-i} A_x^{-2}) W ],
\end{align*}
where we have simplified the last term as 
\begin{align*}
A_z^2 W (R^i)^T (A_{wx}^2)^T = A_z^2 R^{-i} W (A_{wx}^2)^T = A_z^2 R^{-i} A_{wx}^2 W.
\end{align*}
Now we have the following simplifications for the three terms.
\begin{align*}
A_{y+z}^2 R^{-i} A_x^2 & = A_{y+z}^2 A_x^{2^{i+1}} R^{-i}, \\
A_{y+z}^2 R^{-i} A_{wx}^2 & = A_{y+z}^2 A_{wx}^{2^{i+1}} R^{-i}, \\
R^{-i} A_x^{-2} & = A_x^{-2^{i+1}} R^{-i}.
\end{align*}
Applying this back we get
\begin{align*}
& [I_m \mid A_z^2 W] F \\
& = \left[ A_{y+z}^2 A_x^{2^{i+1}} R^{-i} \mid \left( A_{y+z}^2 A_{wx}^{2^{i+1}} + A_x^{-2^{i+1}} \right) R^{-i} W \right] \\
& \equiv \left[ I_m \mid R^i A_x^{-2^{i+1}} A_{y+z}^{-2} \left( A_{y+z}^2 A_{wx}^{2^{i+1}} + A_x^{-2^{i+1}} \right) R^{-i} W \right] \\
& = \left[ I_m \mid R^i \left( A_{w}^{2^{i+1}} + A_{y+z}^{-2} A_x^{-2 \cdot 2^{i+1}} \right) R^{-i} W \right] \\
& = \left[ I_m \mid A_w^2 W + R^i A_{y+z}^{-2} A_x^{-2 \cdot 2^{i+1}} R^{-i} W \right] \\
& = \left[ I_m \mid A_w^2 W + A_{y+z}^{-2^{1-i}} R^i A_x^{-2 \cdot 2^{i+1}} R^{-i} W \right] \\
& = \left[ I_m \mid A_w^2 W + A_{y+z}^{-2^{1-i}} \left( R^i A_{x^{2^{i+1}}}^{-2} R^{-i} \right) W \right] \\
& = \left[ I_m \mid A_w^2 W + A_{y+z}^{-2^{1-i}} A_{x^{2^{i+1}}}^{-2^{1-i}} W \right] \\
& = \left[ I_m \mid \left( A_w^2 + A_{(y+z)^{-2^{-i}} x^{-2}}^2 \right) W \right] \\
& = \left[ I_m \mid A_{w + (y+z)^{-2^{-i}} x^{-2}}^2 W \right].
\end{align*}
Now we define $x \triangleq c, w \triangleq \frac{a}{c}$ and $y \triangleq \left( \frac{d}{c} \right)^{2^i}$.
Then we get
\begin{align*}
w + (y+z)^{-2^{-i}} x^{-2} & = \frac{a}{c} + \left( z + \left( \frac{d}{c} \right)^{2^i} \right)^{-2^{-i}} \frac{1}{c^2} \\
  & = \frac{a}{c} + \left( z^{2^{-i}} + \left( \frac{d}{c} \right)^{2^i \cdot 2^{-i}} \right)^{-1} \frac{1}{c^2} \\
  & = \frac{1}{c} \left[ a + \frac{1}{cz^{2^{-i}} + c \cdot \frac{d}{c}} \right] \\
  & = \frac{1}{c} \left[ \frac{acz^{2^{-i}} + ad + 1}{cz^{2^{-i}} + d} \right] \\
  & = \frac{az^{2^{-i}} + \left( \frac{ad+1}{c} \right)}{cz^{2^{-i}} + d} \\
  & = \frac{az^{2^{-i}} + b}{cz^{2^{-i}} + d} \ ; \ b = \frac{ad + 1}{c}.
\end{align*}
Hence we have proved that $F$ performs the permutation
\begin{align*}
[I_m \mid A_z^2 W] \longmapsto \left[ I_m \mid A_{\frac{az' + b}{cz' + d}}^2 W \right], \ z' \triangleq z^{2^{-i}}.
\end{align*}
We note that the above definitions for $x,w,y$ also satisfy the special case of $i=0$ that corresponds to the proof in Lemma~\ref{lem:P_structure}.
We now substitute these back in $F$ and observe the following simplifications.
\begin{align*}
(A_y^2 R^{-i}) A_x^2 = R^{-i} & A_y^{2^{-i+1}} A_x^2 = R^{-i} A_{y^{2^{-i}} x}^2 = R^{-i} A_d^2, \\
(A_y^2 R^{-i}) A_{wx}^2 + R^{-i} A_x^{-2} & = R^{-i} A_{d/c}^2 A_a^2 + R^{-i} A_{c^{-1}}^2 = R^{-i} A_b^2, \\
W^{-1} R^{-i} A_x^2 & = W^{-1} R^{-i} A_c^2 = (R^i)^T W^{-1} A_c^2, \\
(R^i)^T (A_{wx}^2)^T & = (R^i)^T (A_a^2)^T.
\end{align*}
These imply that the general form of an element in $\mathfrak{P}_{\text{K},m}^*$ is
\begin{align*}
F & = 
\begin{bmatrix} 
A_{y}^2 R^{-i} A_x^2 & (A_{y}^2 R^{-i} A_{wx}^2 + R^{-i} A_x^{-2}) W \\
W^{-1} R^{-i} A_x^2 & (R^i)^T (A_{wx}^2)^T
\end{bmatrix} \\
  & = 
\begin{bmatrix}
R^{-i} A_d^2 & R^{-i} A_b^2 W \\
W^{-1} R^{-i} A_c^2 & (R^i)^T (A_a^2)^T
\end{bmatrix}.   \tag*{\IEEEQEDhere}
\end{align*}

\newpage

\begin{IEEEbiographynophoto}{Trung Can,} a native of Ho Chi Minh City, Vietnam, received a B.A. in Math from Duke University in 2018 and is currently a Ph.D. student at the California Institute of Technology. 

In high school, he received a Gold medal at the International Mathematical Olympiad, representing his home country, Vietnam. At Duke, he served as the vice-president of the Math Union. His undergraduate thesis on the applications of the Heisenberg-Weyl group and finite symplectic geometry received the highest distinction from the math department. He founded and currently serves as the president of PiMA, a nonprofit organization that organizes summer camps and seminars on math topics for high school students in Vietnam. 

His current research interests are algebraic geometry and number theory; in particular, the Schottky problem and Shimura varieties.

\end{IEEEbiographynophoto}

\begin{IEEEbiographynophoto}{Narayanan Rengaswamy} received his Ph.D. in Electrical Engineering from Duke University, USA, in May 2020 where he was supervised by Prof. Henry Pfister and Prof. Robert Calderbank. He is currently a research associate at Duke University continuing to work with his Ph.D. supervisors. His primary training is in classical information and coding theory, and compressed sensing. During his Ph.D. he actively learned about quantum computing and quantum error correction and this is now his primary area of research. Prior to this, he received his Master of Science (M.S.) degree, also in Electrical Engineering, from Texas A\&M University, USA, in December 2015. Earlier he obtained his Bachelor of Technology (B. Tech.) degree in Electronics and Communication Engineering from Amrita University, Coimbatore, India in May 2013, where he ranked first in the college and third in the university.

He has studied the problem of realizing logical operations on quantum error correcting codes in a fault-tolerant manner and produced new insights via a perspective rooted in classical coding theory. His paper on developing conditions for a general stabilizer code to support transversal T gates was one among the 73 out of 283 submissions that were accepted as talks to the esteemed 2020 Quantum Information Processing (QIP) conference. His recent work on Joseph Renes' belief propagation algorithm that passes quantum messages provides a potential near-term application with a quantum advantage for a special-purpose all-photonic quantum processor. In general, he is excited about problems in the intersection of classical and quantum information processing and mathematical tools that help address problems of this sort. He has been a reviewer for IEEE Transactions on Information Theory, IEEE Transactions on Vehicular Technology, Quantum, Quantum Science and Technology, and Proceedings of the Royal Society A, besides the conferences IEEE International Symposium on Information Theory (2018, 2020) and IEEE Information Theory Workshop (2018). He also helped organize the 2016 North American School of Information Theory at Duke University.

\end{IEEEbiographynophoto}

\begin{IEEEbiographynophoto}{Robert Calderbank}(M’89–SM’97–F’98) received the B.S. degree in 1975 from Warwick University, England, the M.S. degree in 1976 from Oxford University, England, and the Ph.D. degree in 1980 from the California Institute of Technology, all in Mathematics. Dr. Calderbank is Professor of Electrical and Computer Engineering at Duke University where he directs the Rhodes Information Initiative at Duke (iiD). Prior to joining Duke in 2010, Dr. Calderbank was Professor of Electrical Engineering and Mathematics at Princeton University. Prior to joining Princeton in 2004, he was Vice President for Research at AT\&T, responsible for directing the first industrial research lab in the world where the primary focus is data at scale. At the start of his career at Bell Labs, innovations by Dr. Calderbank were incorporated in a progression of voiceband modem standards that moved communications practice close to the Shannon limit. Together with Peter Shor and colleagues at AT\&T Labs he developed the mathematical framework for quantum error correction. He is a co-inventor of space-time codes for wireless communication, where correlation of signals across different transmit antennas is the key to reliable transmission.

Dr. Calderbank served as Editor in Chief of the IEEE TRANSACTIONS ON INFORMATION THEORY from 1995 to 1998, and as Associate Editor for Coding Techniques from 1986 to 1989. He was a member of the Board of Governors of the IEEE Information Theory Society from 1991 to 1996 and from 2006 to 2008. Dr. Calderbank was honored by the IEEE Information Theory Prize Paper Award in 1995 for his work on the Z4 linearity of Kerdock and Preparata Codes (joint with A.R. Hammons Jr., P.V. Kumar, N.J.A. Sloane, and P. Sole), and again in 1999 for the invention of space-time codes (joint with V. Tarokh and N. Seshadri). He has received the 2006 IEEE Donald G. Fink Prize Paper Award, the IEEE Millennium Medal, the 2013 IEEE Richard W. Hamming Medal, and the 2015 Shannon Award. He was elected to the US National Academy of Engineering in 2005.

\end{IEEEbiographynophoto}

\begin{IEEEbiographynophoto}{Henry Pfister} received his Ph.D. in Electrical Engineering in 2003 from the University of California, San Diego and is currently a professor in the Electrical and Computer Engineering Department of Duke University with a secondary appointment in Mathematics.  Prior to that, he was an associate professor at Texas A\&M University (2006-2014), a post-doctoral fellow at the {\'E}cole Polytechnique F{\'e}d{\'e}rale de Lausanne (2005-2006), and a senior engineer at Qualcomm Corporate R\&D in San Diego (2003-2004).  His current research interests include information theory, error-correcting codes, quantum computing, and machine learning.

He received the NSF Career Award in 2008 and a Texas A\&M ECE Department Outstanding Professor Award in 2010.  He is a coauthor of the 2007 IEEE COMSOC best paper in Signal Processing and Coding for Data Storage and a coauthor of a 2016 Symposium on the Theory of Computing (STOC) best paper.  He has served the IEEE Information Theory Society as a member of the Board of Governors (2019-2022), an Associate Editor for the IEEE Transactions on Information Theory (2013-2016), and a Distinguished Lecturer (2015-2016).  He was also the General Chair of the 2016 North American School of Information Theory.

\end{IEEEbiographynophoto}

\vfill


\end{document}